\documentclass[useAMS,usenatbib,letterpaper]{mn2e}
\usepackage{graphicx}

\def\ltsima{$\; \buildrel < \over \sim \;$}
\def\gtsima{$\; \buildrel > \over \sim \;$}
\def\simlt{\lower.5ex\hbox{\ltsima}}
\def\simgt{\lower.5ex\hbox{\gtsima}}
 
\title{A model for the metallicity evolution of damped Lyman-$\alpha$
systems}\author[P.H.~Johansson and
G.~Efstathiou]{P.H.~Johansson\thanks{Email: phjohans@ast.cam.ac.uk}
and G.~Efstathiou\thanks{Email: gpe@ast.cam.ac.uk}\\ Institute of
Astronomy, Madingley Road, Cambridge CB3 OHA, UK}
\begin{document}

\date{Version of \today}
\pagerange{\pageref{firstpage}--\pageref{lastpage}} \pubyear{2005}

\maketitle

\label{firstpage}

\begin{abstract}
We apply a physically motivated stellar feedback model to analyse the
statistical properties of damped Lyman-$\alpha$ systems (DLAs)
expected in the concordance cold dark matter (CDM) model.  Our
feedback model produces extended low-metallicity cold gaseous discs
around small galaxies.  Since the space density of galaxies with low
circular speeds is high, these discs dominate the cross-section for
the identification of DLAs at all redshifts. The combined effects of
star formation, outflows and infall in our models result in mild
evolution of the $N_{\rm HI}$-weighted metallicity content in DLAs
with redshift, consistent with observations.  According to our model,
DLAs contribute only a small fraction of the volume averaged star
formation rate at redshifts $z \simlt 5$.  Our model predicts weak
evolution in $\Omega_{\rm HI}$ over the redshift range
$z=0-5$. Furthermore, we show that the cosmological evolution of
$\Omega_{\rm HI}$ and the cosmic star formation rate are largely
disconnected and conclude that the evolution of $\Omega_{\rm HI}$ as a
function of redshift is more likely to tell us about feedback
processes and the evolution of the outer gaseous components of small
galaxies than about the cosmic history of star formation.

\end{abstract}



\section{Introduction}
\label{sec:intro}

Damped Lyman-$\alpha$ systems (DLAs) are defined as quasar absorption
systems with a column density of $N_{ \rm HI} \geq 2 \times 10^{20}
\rm cm^{-2}$, {\it i.e.} with neutral hydrogen column densities
similar to present day galactic discs \citep[see {\it
e.g.}][]{2005MNRAS.364.1467Z}. DLAs probe the high column density end
of the distribution of absorption line systems and are particularly
interesting because their high concentrations of neutral hydrogen
suggests that there may be a close connection with disc galaxies that
we see today \citep[]{1986ApJS...61..249W}. In addition observations
of DLAs measure, in a model independent way, the total neutral
hydrogen mass density of the Universe as a function of redshift.

There are two main competing scenarios for the origin of the DLAs. In the
first model, the DLAs are interpreted as large high-redshift progenitors of
present day massive spiral discs
\citep[]{1986ApJS...61..249W,1991ApJS...77....1L}.  These discs are
assumed to have formed at high ($z\simgt 5$) redshift and to have
evolved little apart from converting their gas into stars. In support
of this theory \citet{1998ApJ...507..113P} argue that the kinematics
of metal absorption lines in DLAs are best explained by models of
rapidly rotating ($v_{c}\simgt 200\;{\rm km}{\rm s}^{-1}$) galactic discs. The
alternative picture, which is more in line with current views of
hierarchical structure formation, posits that the DLA systems are
`dwarf' galaxies rather than fully formed $\sim L^*$ discs 
\citep[]{1994ApJ...430L..97K,1994ApJ...430L..25M}. In particular,
\citet{1998ApJ...495..647H} used hydrodynamic simulations to show that
the statistical properties of the velocity structure seen in the metal
lines could be reproduced by infalling sub-galactic clumps within dark
matter haloes with low virial velocities of $(v_{\rm vir} \simlt 100 \;
{\rm km}{\rm s}^{-1})$.

On the observational side, the nature of the DLA host galaxies 
at high redshift is largely unknown,  with only a handful of optical
detections. \citet{2002ApJ...574...51M} identified three DLA host
galaxies at $z\sim 2-3$ and found them to have luminosities much
fainter than $L^{*}$. At lower redshifts,  $z \simlt 1$,  approximately
$30\%$ of the DLA
host galaxies have luminosities larger than $L^{*}$ \citep[see 
][]{2005MNRAS.364.1467Z}. However, it is dangerous to 
extrapolate from low redshift to high redshift since it is plausible
(perhaps likely) that the cross-section based selection criteria
select  different objects at different redshifts.

The first comprehensive metallicity surveys of DLAs
\citet{1994ApJ...426...79P} derived a global mean $N_{\rm HI}$-weighted
metallicity of $\log(Z/Z_{\odot}) \approx -1$ at $z\approx 2$. Further
observations \citep[]{1999ApJ...510..576P,2000ApJ...533L...5P} found
similar metallicities and no evidence for evolution of the metallicity
in the redshift range $z \sim 1-3.5$. At lower redshift $(z\simlt 1)$ the
situation is more uncertain because of the small sample of DLAs. Recent
measurements at $z \simlt 1$ by \citet{2005ApJ...618...68K} indicate that the
evolution of metallicity for DLAs remains weak at lower redshifts and  does
not rise up to solar, or near-solar, values by $z=0$. In
addition, the cosmological density of neutral
hydrogen ($\Omega_{\rm HI}$) as traced by the DLA population shows little
evolution between $z\approx 0.5- 5$ \citep{2006ApJ...636..610R}.

Attempts to model DLAs in cosmological simulations have proved
difficult because of the high numerical resolution required and the
difficulties in modelling stellar feedback and galactic
winds\citep[see {\it
e.g.}][]{1996ApJ...457L..57K,1997ApJ...484...31G,2001ApJ...559..131G}. In
addition to limited numerical resolution, early simulations did not
follow the chemical evolution of the DLAs in any detail. Some of these
shortcomings have been overcome with recent simulations by
\citet{2003ApJ...598..741C} and \citet{2004MNRAS.348..435N}. Both
studies are able to reproduce the flat metallicity evolution of the
DLAs, but their metallicity values tend to lie at around
$\log(Z/Z_{\odot}) \approx -0.5$, considerably lower than observed.
To reconcile this higher value with the observations, both groups
invoke a large dust obscuration correction. A dust bias has also been
invoked to explain the lack of DLA absorbers with $[ \rm Zn/H]+ \log
[N( \rm HI)] > 21$ noted by \citet{1998A&A...333..841B}.  However, a
recent study by \citet{2004MNRAS.354L..31M} using a large sample of
quasar spectra from the SDSS Data Release 2 did not find any evidence
for dust-reddening caused by intervening DLAs at $z\sim 3$.  In
addition DLAs extracted from samples of radio selected QSOs, which
should be largely free from any dust bias, show only marginally higher
metallicities compared to absorbers from optically selected control
samples
\citep[]{2001A&A...379..393E,2005A&A...440..499A,2006ApJ...646..730J}.

Semi-analytic modelling of DLAs provides a complementary approach to
numerical simulations
\citep[]{1996MNRAS.281..475K,2000MNRAS.315...82P,2001MNRAS.326.1475M,2001MNRAS.320..504S,2004ApJ...603...12O}. 
The semi-analytic approach is, of course, much
faster than numerical simulations and so can be used to explore
parametric representations of physically complex processes such as
stellar feedback. Semi-analytic modelling can also provide physical
insight, which is sometimes difficult to acquire from
numerical simulations. The disadvantage of semi-analytic modelling is
that some aspects of the models may be sensitive to processes that are
extremely difficult to model analytically. Even if this is the case
(as it is in the models described in this paper) it is useful to
identify aspects of a theoretical model that are robust, and can be
understood analytically, and those that are likely to require
further investigation using numerical simulations.

In this paper we develop a semi-analytic model of DLAs using the
feedback model described by \citet{2000MNRAS.317..697E} (hereafter
E00). Some aspects of the application of this feedback model to DLAs
were sketched in E00. The purpose of this paper is to develop a more
detailed model that can be compared with a wide range of new
observations.  The E00 feedback model produces extended cold gaseous
discs around `dwarf' galaxies (defined crudely in this paper as any
system in a halo with a circular speed $v_c \simlt 100\; {\rm
km} {\rm s}^{-1}$) .  In CDM-like models such gaseous discs would dominate
the cross-section for the identification of DLAs at high redshift
because the space density of haloes with low circular velocity is high
\citep[]{1994ApJ...430L..97K,1994ApJ...430L..25M}. In our model, most
of the cross-section is dominated by largely unprocessed gas in the
outer parts of the galaxies that does not participate in the bulk of
the star formation process. As a result, the evolution of $\Omega_{\rm
HI}$ in this model is linked only indirectly to the star formation
history.  There are some similarities between our models and recent
semi-analytic modelling of DLAs by \citet{2004ApJ...603...12O}. The
main differences, which are particularly significant for modelling the
metallicities of DLAs, are: (i) our stellar feedback and outflow
models are physically motivated, {\it i.e.} they are related to a well
defined, though simplified, model of the interstellar medium in
galaxies; (ii) rather than applying ad-hoc rules of star formation and
chemical enrichment, our star formation prescription is based on the
self-regulated\footnote{There is considerable empirical support in
favour of self-regulated star formation in disc systems
\citep{1998ARAA..36..189K}.}  model described in E00.

The structure of the paper is as follows.  Section \ref{sec:feedback}
summarizes the feedback model and its application to DLAs.  In Section
\ref{sec:chemenrich} the feedback model is applied to calculate the
metallicity evolution of the DLA population. Theoretical predictions
of various global properties of the DLA population are described in
Section \ref{sec:globprop}. In Section \ref{sec:conc} we summarize our
results and present our conclusions.

\begin{figure*}
 \begin{center}
 {\includegraphics[scale=0.8]{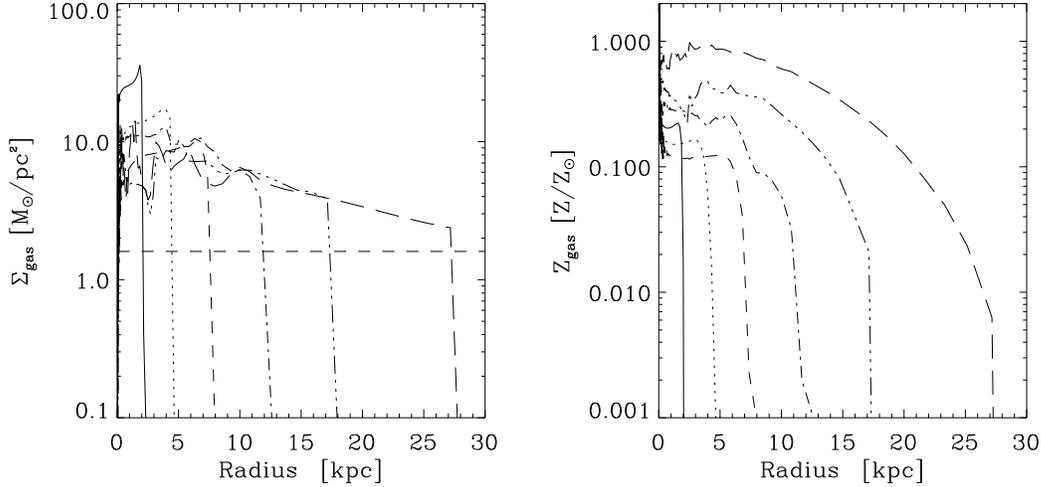}}
 \end{center}
 \caption[]{The left-hand panel shows the evolution of hydrogen gas
            surface density and the right-hand panel the evolution of
            the gas metallicity for a model with $\lambda=0.075,
            v_{\rm vir}=100\; {\rm km}{\rm s}^{-1}$. The results are plotted 
            at ages of 0.4 Gyr (solid line), 0.8 Gyr (dotted line), 1.4 Gyr 
            (dashed line), 2.6 Gyr (dash-dotted line), 5.2 Gyr (dash-triple dotted line) 
	    and 10 Gyr (long dashed line). 
	    The horizontal dotted line in the left-hand panel shows the DLA selection limit 
            of $1.6 \ M_{\odot} \rm pc^{-2}$.}
 \label{fig:Rad_gasmet_prof}
\end{figure*}


\section{The model}
\label{sec:feedback}

The E00 feedback model is used to evolve a grid of galaxy models as
 described below.  In this Section we will summarize briefly some
 aspects of the model and describe the relevant parameters
 involved. We refer the reader to E00 for a more detailed description
 of the model. All calculations in this paper assume a concordance CDM
 cosmology with parameters\footnote{where $\Omega_{m}$,
 $\Omega_{\Lambda}$ and $\Omega_{b}$ are the cosmological densities of
 matter, dark energy and baryons respectively and $h$ is defined such
 that $H_{0}$=100$h$ kms$^{-1}$Mpc$^{-1}$} $\Omega_{m}=0.27$, $\Omega_{\Lambda}=0.73$,
 $\Omega_{b}=0.044$, $h=0.71$, {\it e.g.}
 \citep{2003ApJS..148..175S}. We assume a scale-invariant adiabatic
 fluctuation spectrum, normalized so that $\sigma_{8}=0.84$.

\subsection{Feedback model}

E00 developed a self-regulated feedback model in which the
instantaneous star formation rate is regulated by disc instabilities
and in which supernovae blastwaves convert some fraction of the cold
disc gas into a hot component which can flow out of the system. The
model is based, in part, on the \citet{1977ApJ...218..148M} theory of
the interstellar medium, but also includes the infall of gas within a
dark halo and the formation of a stellar disc through self-regulated
star formation.

The dark matter halo of the model galaxies is described by a static
NFW profile \citep{1996ApJ...462..563N}.  A galaxy forms through
infall of gas from the surrounding halo under the assumption that the
angular momentum of the disc material is approximately conserved
during the collapse of the disc \citep{1980MNRAS.193..189F}.  
The angular momentum distribution within a dark matter halo is based
on a fit to the results of $N$-body simulations and is normalised to
reproduce a target value of the 
dimensionless spin parameter $\lambda_{H}=J|E|^{1/2}G^{-1}M^{-5/2}$,
where $J$, $E$ and $M$ are the angular momentum, binding energy,
and mass within the virial radius  $r_{V}$.

The stability of the resulting gaseous disc is determined by a
\citet{1965MNRAS.130..125G} like criterion,
\begin{equation}
  \sigma_{g}=\frac{\pi G \mu_{g}}{\kappa} g(\alpha,\beta),
\label{eq:stability} 
\end{equation}
where $\mu_{g}$ is the gas surface density, $\kappa$ is the epicyclic frequency, $\alpha$ and $\beta$ are defined 
as $\sigma_{\star}=\alpha\sigma_{g}$, $\mu_{\star}=\beta\mu_{g}$ and $g(\alpha,\beta)$ is a correction factor
of order unity that applies to a two-component gaseous-stellar disc (see E00 for details and Fig 1. in E00
for a plot of $g(\alpha,\beta)$ as a function of $\alpha$ and $\beta$).

Following \citet{1977ApJ...218..148M} the interstellar medium of the
galaxy is modelled as a multiphase medium.  Most of the gas is assumed
to be in cold clouds. The system of cold clouds is assumed to be
marginally unstable and so their velocity dispersion is fixed in terms
of the surface mass densities of gas and stars by equation
(\ref{eq:stability}).  Supernova blastwaves propagating through the
interstellar medium convert some of the cold clouds into a low-density
hot component. The instantaneous star formation rate is fixed by
balancing the energy dissipated in collisions between cold clouds with
that supplied by supernovae shells.

Assuming a standard \citet{1955ApJ...121..161S} stellar initial mass
function (IMF) with mass cutoffs of $m_{\rm l}=0.1 M_{\odot}$ and
$m_{\rm u}=50 M_{\odot}$ and that each star of mass greater than $8
M_{\odot}$ releases $10^{51} E_{51} \ \rm erg$ in kinetic energy in a
supernova explosion, the energy injection rate is related to the star
formation rate by
\begin{equation}
\dot{E}_{\rm SN}=2.5 \times 10^{41}  E_{51} \dot{M_{\star}} \rm \ erg \ s^{-1},
\label{eq:sn} 
\end{equation} 
where $\dot{M_{\star}} $ is the star formation rate in $M_{\odot}$ per year.
A fraction of this energy rate, $\epsilon_{c} \dot{E}_{SN}$ is assumed to balance the
rate of energy loss per unit surface area due to cloud collisions 
\begin{equation}
\dot{E}^{\Omega}_{\rm coll}=5.0 \times 10^{29}\left(1+\frac{\beta}{\alpha} \right) \mu^{3}_{g5}\sigma_{g5} \rm \ erg \ s^{-1} pc^{-2}, 
\label{eq:coll} 
\end{equation} 
where $\mu_{g5}$ is the surface mass density of the gas component in units of $5 M_{\odot}\rm pc^{-2}$ and $\sigma_{g5}$ 
is the cloud velocity dispersion in units of $5 \rm \ kms^{-1}$.

The uncertain parameter $\epsilon_{c}$ is fixed by normalizing the
 typical star formation rate for a Milky Way type galaxy.  Assuming a
 flat surface mass density profile for the gas to $R_{\rm max}=14 \ \rm
 kpc$ and $\beta \approx10$, $\alpha \approx 5$ results in
\begin{equation}
\epsilon_{c}\dot{M_{\star}}=0.004.
\label{eq:SF} 
\end{equation}
The choice $\epsilon_{c}=0.01$ results in a reasonable net star
 formation rate of $0.4 \ M_{\odot}\rm yr^{-1}$ for a Milky Way-type
 galaxy. We therefore adopt $\epsilon_{c}=0.01$ throughout this
 paper. We also assume a cooling function for gas of a primordial
 composition with a sharp lower cutoff at $T=10^{4} \rm K$.

Following \citet{1977ApJ...218..148M} an expanding supernova remnant will evaporate a mass of 
\begin{equation}
M_{\rm ev} \approx 311 E_{51}^{6/5}(4 \pi a_{\rm l}N_{\rm cl}\phi_{\kappa})^{3/5} n^{-4/5}_{\rm h} M_{\odot}
\label{eq:massEV} 
\end{equation}
where $n_{\rm h}$ is the density interior to the supernova remnant (in
units of $\rm cm^{-2}$), $a_{l\rm }=0.5 \ \rm pc$ is the lower limit
of the distribution of cloud radii and $N_{c\rm l}$ is the number
density of the clouds (in units of $\rm pc^{-3}$). (Note that for the typical sizes and 
densities of the cold clouds, the filling factor for HI absorption will always be greater
than unity for HI column densities above the DLA threshold).
The second free parameter of the model, $\phi_{\kappa}$, quantifies the effectiveness
of classical thermal evaporation (via conductivity) of the clouds
$\kappa_{\rm eff}=\kappa\phi_{\kappa}$ and is expected to be less than
unity if conductivity is reduced by tangled magnetic fields,
turbulence, etc. The parameter $\phi_{\kappa}$ controls the strength
of stellar feedback.  In our normal model we adopt
$\phi_{\kappa}=0.1$, but we have also run models with weaker feedback
$\phi_{\kappa}=0.01$ and stronger feedback $\phi_{\kappa}=1.0$, (WFB
and SFB models, respectively) so that the reader can gauge the
sensitivity of our results to uncertainties in the strength of stellar
feedback. 

The cold gas component is assumed to be in the form of HI, which is clearly an
oversimplification since some of this gas could be converted to molecular hydrogen. 
However, this is likely to occur in the central high-density regions of galaxies
rather than the outer parts that dominate the DLA cross-section. 
This together with the low molecular gas fractions observed for
DLA systems (see \citealt{2005ARA&A..43..861W} and references therein) suggests that 
ignoring the molecular component will not affect our results. We also ignore the 
metagalactic UV flux since at the column densities of the DLA systems they are of 
course self-shielded to photoionization.

The hot phase of the interstellar medium can escape the galaxy if the
wind speed $v_{\rm w}$ exceeds the escape velocity $v_{\rm esc}$ from
the centre of the galaxy (see E00 for details on  how $v_{\rm w}$ is
calculated). A wind with $v_{\rm w}<v_{\rm esc}$ is assumed to
participate in a galactic fountain and is therefore returned to the
disc on a dynamical timescale, as described in E00. The model thus
incorporates simultaneous inflow and outflow of gas ignoring any
interactions between these flows. As discussed in E00 this may not
be a bad approximation since the feedback may produce a mildly collimated
outflow with infall occurring primarily in the equatorial plane, this type of behaviour has
been seen in numerical simulations with stellar feedback \citep[see {\it
e.g.}][]{2003ApJ...587...13T,2003MNRAS.339..289S}

Chemical evolution is included using the instantaneous recycling
approximation. Separating between primordial infalling gas accreting
at a rate $d\mu_{I}/dt$ with metallicity $Z_{I}$ and processed gas
from the galactic fountain of metallicity $Z_{F}$ accreted at a rate
$d\mu_{F}/dt$ results in
\begin{equation}
\mu_{g}dZ=pd\mu_{S}+(Z_{I}-Z)d\mu_{I}+(Z_{F}-Z)d\mu_{F}.
\label{eq:chemev} 
\end{equation}
Here $p=0.02$ \citep[]{2005A&A...433.1013H} is the yield and, unless
otherwise stated, we assume that primordial gas has zero
metallicity. The metallicities are normalized to the solar value for
which we adopt the new value of $Z_{\odot}=0.0133$ by
\citet{2003ApJ...591.1220L}.

 \begin{figure*}
 \begin{center}
   {\includegraphics[scale=0.8]{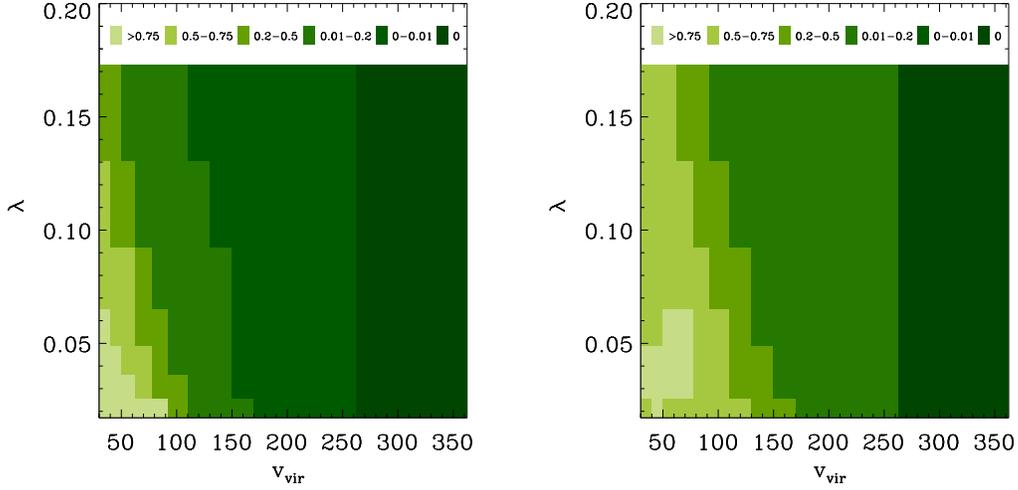}}
 \end{center}
 \caption[]{The ejection fractions defined as $f_{\rm ej}=M_{\rm ej}/M_{\rm tot}$, where $M_{\rm tot}=M_{\rm ej}+M_{\rm gas}+M_{\rm star}$ 
            is the total baryonic mass. The results are plotted at $z=0$ (left panel) and $z=3$ (right panel)
            as a function of the spin parameter $\lambda$ and the virial velocity of the halo $v_{\rm vir}$ for the 119 models. 
            The ejection fraction shows a strong increase with decreasing $\lambda$ and $v_{\rm vir}$.}
 \label{fig:Eject}
\end{figure*}

 \begin{figure*}
 \begin{center}
   {\includegraphics[scale=0.8]{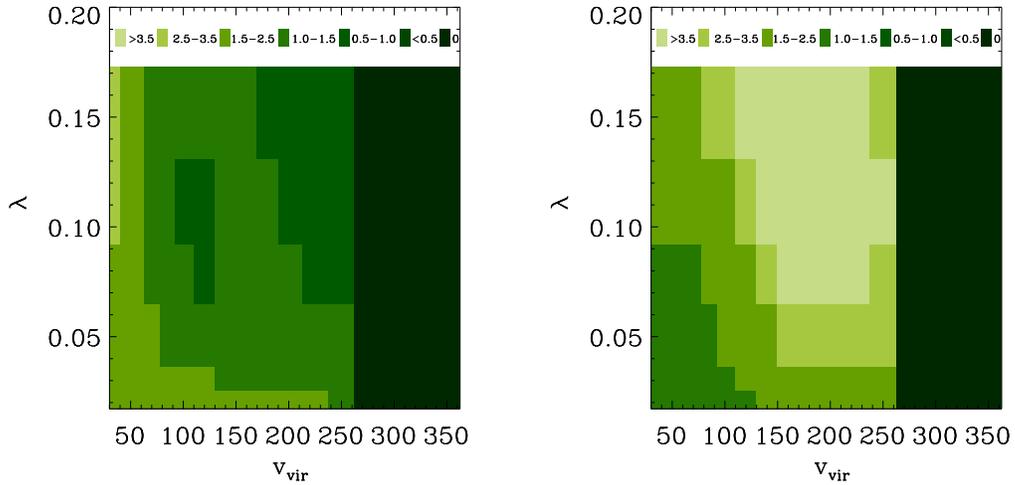}}
 \end{center}
 \caption[]{The metallicity of the ejected gas relative to the galaxy gas metallicity $(Z_{\rm ej}/Z_{\rm g})$ at $z=0$ (left panel) 
and $z=3$ (right panel) as a function of the spin parameter $\lambda$ and the virial velocity of the halo $v_{\rm vir}$ for the 119 models. The 
ejected gas is predominantly enriched relative to the gas in the parent galaxy.}
 \label{fig:Meteject}
\end{figure*}

\subsection{Modelling the DLA population}
\label{sec:Modelling}

Using the model described above we computed the gas and stellar
surface densities, and their respective metallicity profiles, for a
grid of models parameterized by the virial velocity of the parent dark
halo (defined as the circular speed at the virial radius) and the spin
parameter $\lambda$.  We computed a grid of 17 models with virial
velocities in the range $v_{\rm vir}=35-350 \ \rm km/s$ and 7 spin
parameters in the range $\lambda=0.021-0.15$ resulting in a total of
119 models. The lower cutoff of the virial velocity was chosen because
gas in haloes with low circular speeds will be heated by the external
UV background at high redshifts and so is unable to cool
\citep[]{1986MNRAS.218P..25R,1992MNRAS.256P..43E}. (The precise value
of this lower cutoff is unimportant for most of the results presented
in this paper. Over the redshift range $0 \simlt z \simlt 5$, the
smaller cross-section of such low circular speed systems more than
compensates for their larger space density. As discussed in Section 4,
the peak contribution to the DLA population occurs at higher circular
speeds.)  Radial profiles for each model were stored for thirteen output
times in the range $0.17-13.5 \ \rm Gyr$

Following \citet[]{1993MNRAS.262..627L} we define the formation time
of a halo as the time when it has assembled half of its final mass.
The distribution of formation redshifts for a halo with final mass
$M_{1}$ at redshift $z_{1}$ can therefore be approximated by
\begin{equation}
p(z_{\rm f})dz_{\rm f}=p(\omega)d\omega=2\omega \ \rm erfc \left( \frac{\omega}{\sqrt{2}} \right) d\omega,
\label{eq:formage} 
\end{equation}
where $\omega^{2}=(\delta_{\rm cf}-\delta_{\rm cl})^{2}/(S_{\rm
f}-S_{1})$, $\delta_{\rm cf}=\delta_{\rm c}(z_{\rm f})$, $\delta_{\rm
cl}=\delta_{\rm c}(z_{1})$ and $S_{\rm f}=\sigma^{2}(M_{1}/2)$.  Here
$\delta_{\rm c}(z)$ is the overdensity required for spherical collapse
at redshift $z$ extrapolated using linear theory to the present time and
$\sigma^{2}(M)$ is the variance of the initial density fluctuation
field. Using this description we can estimate (approximately) the age
distribution of haloes as a function of $v_{\rm vir}$ at each output
redshift. At high redshifts the halo age distribution is of course
narrow and broadens significantly at lower redshifts where a large
number of output ages are needed to model the halo population.

At each output redshift we calculated the number density of haloes as
a function of virial velocity using the \citet{1999MNRAS.308..119S}
mass function
\begin{eqnarray}
n_{ST}(M)dM&=&-\left(\frac{2}{\pi}\right)^{1/2} A
\left[1+\left(\frac{a\delta_{c}^{2}}{\sigma^{2}}\right)^{-p}\right]a^{1/2}
\frac{\rho_{b}}{M} \nonumber\\ &&\times
\frac{\delta_{c}}{\sigma^{2}}\frac{d\sigma}{dM}\exp\left(-\frac{a\delta^{2}_{c}}{2\sigma^{2}}\right)
dM
\label{eq:ST} 
\end{eqnarray}
where $A=0.322$, $p=0.3$, $a=0.707$, $\rho_{b}$ is the background
density, $\sigma$ is the rms density fluctuations on scale $R$
corresponding to mass $M$ $(M=4 \pi \rho_{b}R^{3}/3)$ and
$\delta_{c}\approx 1.68$.  

The distribution of halo spin parameters is assumed to follow a
lognormal distribution and is assumed to be independent of both time
and $v_{\rm vir}$
\begin{equation}
p(\lambda)d\lambda=\frac{1}{\sqrt{2\pi}\sigma_{\lambda}}\exp\left[-\frac{\log^{2}(\lambda/\bar{\lambda})}{2 \sigma^{2}_{\lambda}}\right]\frac{d\lambda}{\lambda}
\label{eq:lambda} 
\end{equation}
with $\bar{\lambda}=0.045$ and $\sigma_{\lambda}=0.56$
\citep{2002ApJ...581..799V}.

All haloes are assumed to evolve in isolation. We have not included an
elaborate merger tree to describe the merger history of an individual
dark matter halo and its baryonic content. Our model is thus not
designed to give a description of the history of an individual
galaxy. Instead our model has been constructed to provide an
approximate `snapshot' of the galaxy distribution at each output
redshift characterized by the parameters $(t_{\rm f}, v_{\rm vir},
\lambda)$. By using the simplified model of equations
(\ref{eq:formage}) and (\ref{eq:ST}) we are, in effect, assuming that
the baryonic parts of galaxies follow a similar merger history to that
of the dark matter within the virial radius. This is, of course, a
simplification and one would expect the high density baryonic
components to survive as distinct systems for a longer time than their
parent haloes. Our main results on the metallicity distributions of
DLAs (Section 3) are insensitive to the merger history.  However, some
properties, in particular the numbers of DLAs at high redshift, are
extremely sensitive to the merger history (see Section 4). Our model
can give only a very approximate description of properties that are
sensitive to merger history and further progress will probably require
high resolution gas dynamical numerical simulations.

At each output redshift we calculate properties of the galaxy
distribution along sight-lines assuming galaxies are uniformly
distributed and have random inclination angles. We assumed a hydrogen
fraction of $X=0.7$ by mass. If the probing sight-line encounters a
hydrogen column density of $N_{\rm HI} \geq 2 \times 10^{20} \rm
cm^{-2} \approx 1.6 \ M_{\odot} \rm pc^{-2}$ we select the sightline as a
DLA and measure its physical properties such as column density,
metallicity and star formation rate at the intercept point.

\subsection{Model properties}

In our model, galaxies form in an intense starburst followed by a
rapid buildup of metallicity,  as can be seen in
Fig. \ref{fig:Rad_gasmet_prof} which shows the formation of a disc in
halo with a relative high $\lambda$ and a virial velocity $v_{\rm vir}
= 100\; {\rm km}{\rm s}^{-1}$ (see also E00 for a more extensive
discussion). The initial burst is followed by quiescent star formation
that builds up the galactic disc from inside-out.  The model develops
an extended gaseous disc with surface density well above the DLA
definition of $N_{\rm HI} \geq 2 \times 10^{20} \rm cm^{-2}$. The sharp
outer edge of the gas disc seen in Fig. \ref{fig:Rad_gasmet_prof} is a consequence
of the infall model in which the final time of the model sets the maximum cooling
radius within the halo. The oscillatory behaviour of the radial gas profile
near the centre is a consequence of the galactic fountain and similar
behaviour can be also seen in the star formation profile.

In addition to gaseous infall and star formation, the evolution of
galaxies according to this model is strongly influenced by outflows produced by
stellar feedback.  The ejected mass fraction, $f_{\rm ej}$, at time
$t$, is defined as the ratio of the ejected baryonic mass $M_{\rm ej}$
divided by the total baryonic mass $M_{\rm tot}=M_{\rm ej}+M_{\rm
gas}+M_{\rm star}$, where $M_{\rm gas}$ and $M_{\rm star}$ are the
mass in the cold gaseous disc and the stellar disc at time $t$. The
parameter $f_{\rm ej}$ is a strong function of both $v_{\rm vir}$ and
$\lambda$ as can be seen in Fig. \ref{fig:Eject}. According to our
model,  galaxies with small $v_{\rm vir}$ and $\lambda$ can have 75\% or
more of their gas mass ejected by $z=0$. This is expected to have a
large effect on various properties (particularly metallicities) of DLAs as
the low mass haloes dominate the cross-section especially at high
redshift (see Fig. \ref{fig:mean_halo_prop}).

In addition to their large ejection fraction, galaxies with small
$v_{\rm vir}$ and $\lambda$ predominantly eject gas that is enriched
relative to the host galaxy.  This is especially true at high
redshifts as can be seen in the right-hand plot of
Fig. \ref{fig:Meteject}. By $z=0$ the metallicity enhancement of the
ejected material is lower because most of the ejection occurs at high
redshift and the galactic disc, by $z=0$, has had time to evolve to
higher metallicity.

The effectiveness of supernova feedback in ejecting gas from 
spiral and irregular galaxies has also been studied observationally by \cite{2002ApJ...581.1019G}. They found 
an increasing correlation for both the metallicity $Z$ and the effective
yield $y_{\rm eff}$ as a function of rotational velocity $v_{\rm rot}$. However, at 
$v_{\rm rot}\sim 150  \ \rm km/s$ both correlations turn over and both
the $Z$ and $y_{\rm eff}$ become approximately constant for 
$v_{\rm rot}\simgt 150  \ \rm km/s$. \cite{2002ApJ...581.1019G} conclude that this observed trend  
suggests that galaxies with $v_{\rm rot}\simlt 100-150  \ \rm km/s$ may lose 
a large fraction of their metal-enriched gas through supernova feedback, whereas more massive
galaxies tend to retain their metals.

To test this observational picture we calculated the average
stellar metallicities of our model galaxies as a function of $v_{\rm vir}$ and $\lambda$ at different
redshifts. Our models produce an increasing correlation of the stellar metallicity with $v_{\rm vir}$ until a critical $v_{\rm vir}$
is reached, after which the metallicity becomes approximately constant. At $z=0$ this critical velocity
is $v_{\rm vir}\sim 70-85\ \rm km/s$, which corresponds to $v_{\rm rot}\simlt 170-200  \ \rm km/s$ 
for a typical model. Thus the chemical evolution of the stellar component in our model seems to be 
in very good agreement with observations.


\begin{figure*}
 \begin{center}
   {\includegraphics[scale=0.8]{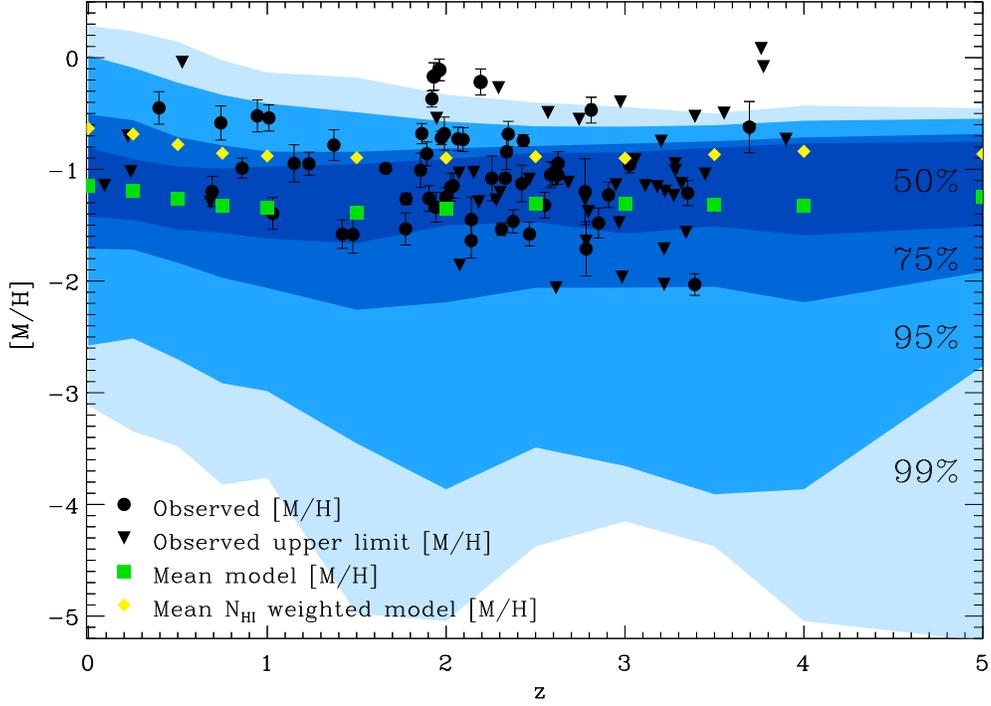}}
   \end{center} \caption[]{The redshift evolution of DLA metallicity
   for our normal feedback model ($\phi_{\kappa}=0.1$) overplotted
   with observations from
   \citep[]{2005A&A...440..499A,2005ApJ...618...68K} and references
   therein.  The (blue) shaded contours give the range of metallicity
   for DLAs at each output redshift. The square (green) points show
   the mean DLA metallicity and diamond (yellow) points show the
   $N_{\rm HI}$-weighted mean metallicity.}
   \label{fig:Met_redshift_norm}
\end{figure*}

\begin{figure*}
 \begin{center}
   {\includegraphics[scale=0.7]{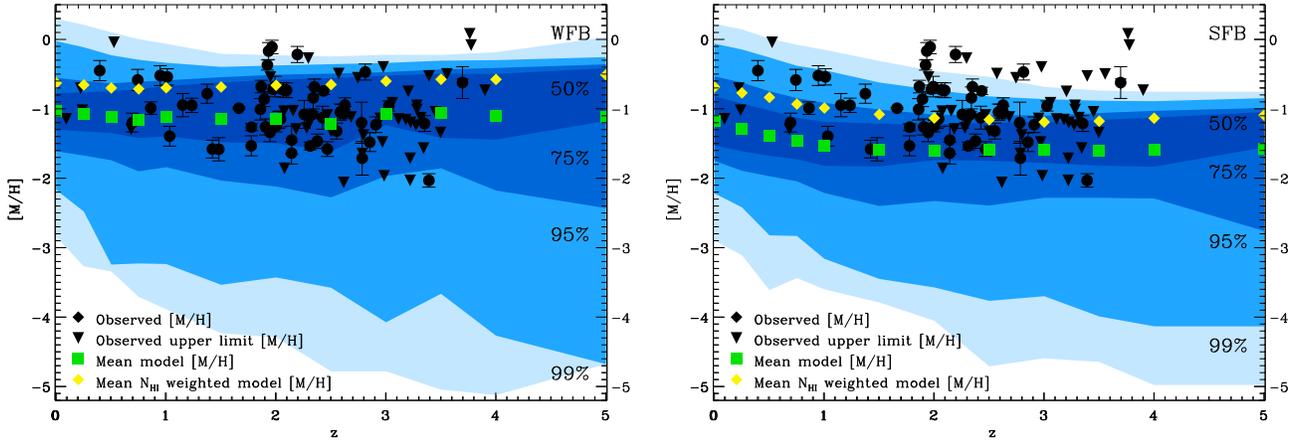}}
   \end{center} \caption[]{The redshift evolution of DLA metallicity
   for our weak (left-hand panel) feedback model
   ($\phi_{\kappa}=0.01$) and strong (right-hand panel) feedback model
   ($\phi_{\kappa}=1.0$) overplotted with the same observational data
   and symbols as Fig. \ref{fig:Met_redshift_norm}.}
   \label{fig:Met_sfb_wfb_redshift_norm}
\end{figure*}

\section{Metallicity evolution of DLAs}
\label{sec:chemenrich}

\subsection{Normal feedback (NFB) model}

The metallicity evolution of the DLAs as a function of redshift is
shown in Fig. \ref{fig:Met_redshift_norm} for the normal feedback
model with $\phi_{\kappa}=0.1$. The square (green) points show the
mean DLA metallicity and the (blue) shaded contours illustrate the 
metallicity distributions  at each output redshift. The diamond (yellow)
points show the predicted $N_{\rm HI}$-weighted mean metallicities.
The metallicity evolution binned in unit redshift bins 
for the observational data set and our feedback model predictions is  
summarized in Table \ref{met_evolution}.

\begin{table*}

 \label{met_evolution}
 \centering
  \begin{tabular}{@{}lcccc}
  \hline
   Data set & $0<z<0.5$ &  $0.5<z<1.5$ & $1.5<z<2.5$  & $2.5<z<3.5$ \\ 
   \hline\hline
   All obs    &     $-0.87 \pm 0.25$ & $-1.02\pm0.27$ & $-1.00 \pm 0.15$ & $-1.20\pm0.18$ \\
   Errbars & $------$ & $-1.06\pm0.28$ & $-1.01 \pm 0.15$ & $-1.17\pm0.28$  \\
   NFB     & $-0.70$  & $-0.88$ &   $-0.90$ & $-0.90$ \\ 
   WFB   & $-0.66$  & $-0.70$ &   $-0.66$ & $-0.60$ \\
   SFB   & $-0.77$  & $-0.99$ &   $-1.13$ & $-1.19$ \\
\hline
\end{tabular}
\caption{The metallicities as a function of redshift for the
observations and for our three feedback model predictions.  `All obs'
includes all observations, with upper limits as if they were
detections, whereas `Errbars' excludes the upper limit data
points. The observations have been included with 95\% uncertainties
given by a bootstrap analysis. For the lowest redshift bin we only
have one observation with errorbars.}
\end{table*}

\normalsize

Figure \ref{fig:Met_redshift_norm} shows observational measurements of
[Zn/H] for DLA systems\footnote{ The observational data points were
kindly provided by Chris Akerman}.  The observational sample consists
of 87 abundance measurements compiled by \citet{2005ApJ...618...68K}
(see references therein) and 16 measurements by
\citet{2005A&A...440..499A} totalling 103 abundance
measurements. These data consist of both observations with error bars
and upper limits indicated by triangles.  We used the [Zn/H] abundance
as an indicator of total metallicity to compare with our models.
 The assumption that zinc is undepleted on grains and traces
the iron abundance over three orders of magnitude in [Fe/H] \citep[see
e.g.][]{1991A&A...246..354S} is strongly supported by the recent work
of \citet{2004A&A...415..993N}. Zinc should therefore provide an
accurate overall indicator of the total metallicity in the gas. In
addition, by using [Zn/H] abundances we can compare our results to a
large observational sample.

The most important result from Fig. \ref{fig:Met_redshift_norm} is that
our model predicts a typical DLA metallicity of $\rm [M/H]\approx
-1.0$ with very little evolution over the entire redshift range $z=5$
to $z=0$. In more detail, the model predicts a mean $N_{\rm
HI}$-weighted metallicity of $\rm [M/H]=-0.86$ at $z=5$ and shows a
slight decrease from $z=5$ to $z\approx2$.  This decrease can be
seen in the model shown in Fig. \ref{fig:Rad_gasmet_prof} and is
accentuated by the stronger outflows of enriched gas from galaxies
with small $v_{\rm vir}\sim 70 \ \rm km/s$ which dominate the DLA
cross-section at high redshift. From $z=2$ to $z=0$ the model shows a
small but steady increase in the average metallicity. This is caused
by two effects: a) galaxies at lower redshifts have had time to build
up higher metallicities via quiescent star formation; b) the DLA
cross-section becomes dominated by galaxies with larger virial
velocities $(v_{\rm vir}=70-100 \ \rm km/s)$ for which the effects of
metal enriched outflows are not as pronounced as for galaxies with
lower $v_{\rm vir}$.  The combined effect of a shift in the typical
$v_{\rm vir}$ of the DLA population with redshift and the dependence
of the metallicity of the outflows as a function of $v_{\rm vir}$ and
time results in almost no evolution in the DLA metallicity. It is
worth emphasising that this follows with no adjustment of the feedback
parameters compared to those used by E00. The parameters of the model
have not been tuned specifically to match the observations plotted
in Fig. \ref{fig:Met_redshift_norm}, instead we adopted a model that roughly matches
 the gross properties of the ISM of the Milky Way and
reproduces its net star formation rate.

\begin{figure*}
 \begin{center}
   {\includegraphics[scale=0.65]{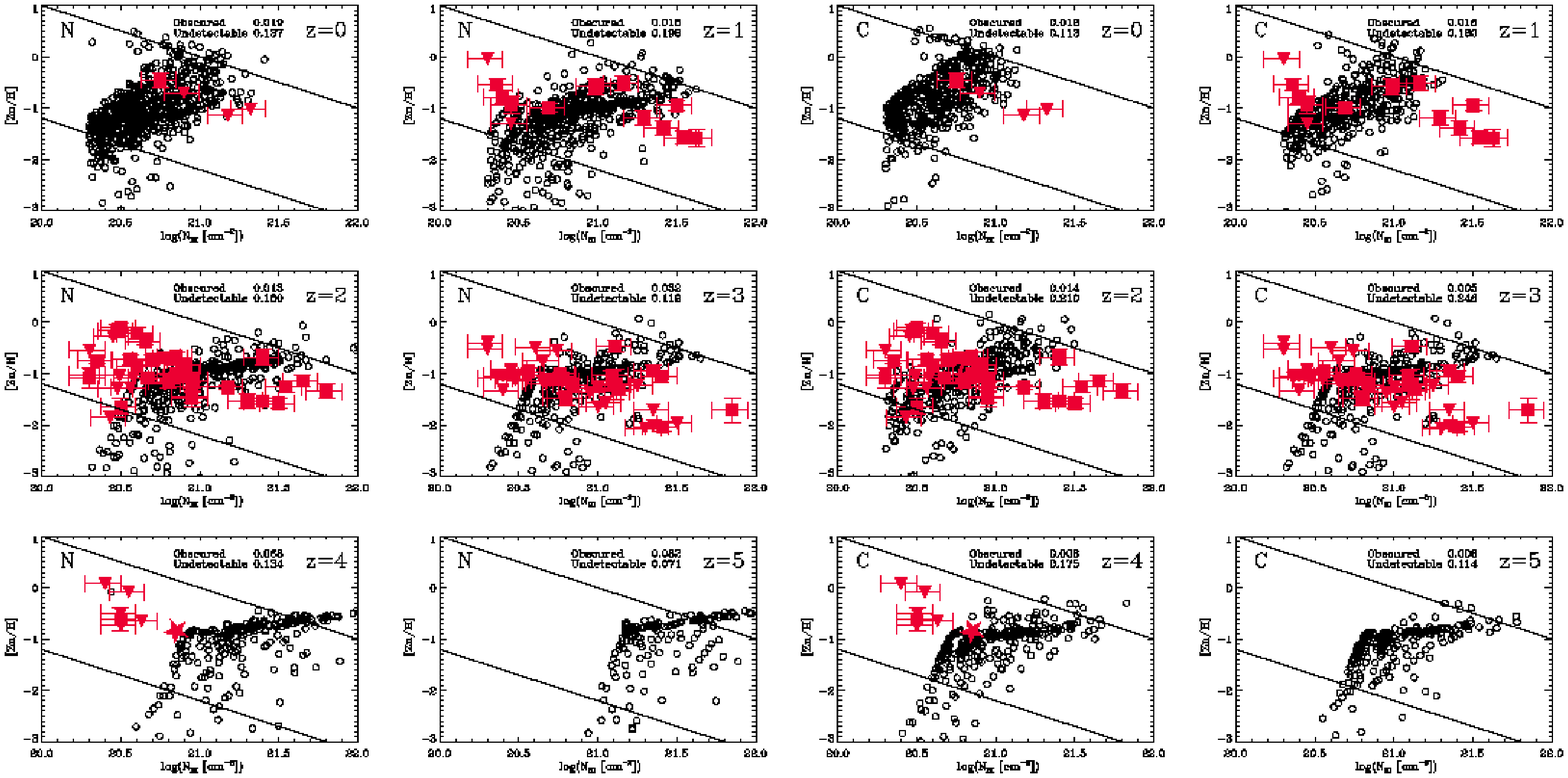}}
 \end{center}
 \caption[]{The HI column density versus gas metallicity at redshifts z=0-5 for the NFB model (left-hand panel) 
	    marked with N and for the NFBc coeval evolution model (right-hand panel) marked with C.
            The model (open circles) is overplotted
            with observations with error bars (red squares) and upperlimits (red triangles) 
            from \citep[]{2005A&A...440..499A,2005ApJ...618...68K} and references therein. In addition we include
            the recent gamma-ray burst DLA (red star) observed by \citet{2006A&A...451L..47F}.
            The observational data has been binned in unit redshift bins at each redshift.} 
 \label{fig:Met_Nh_bias}
\end{figure*}


Performing a least-squares linear fit to the mean $N_{\rm
HI}$-weighted model predictions over the entire redshift range we
derive a slope of $m= -0.04 \ \rm dex/\Delta z$ with a zero point of
$\rm [M/H]_{0}=-0.76 \ \rm dex$ confirming the visual impression of no
evolution within the error bars. The fit to the unweighted mean
metallicities yields $m= -0.02\ \rm dex/\Delta z$ with $\rm
[M/H]_{0}=-1.26\ \rm dex$. The unweighted metallicity tracks the
$N_{\rm HI}$-weighted metallicity accurately with an offset of 0.5
dex.

However the model predicts some evolution in the metallicity
distributions of the DLAs with a tail extending to low metallicities
at redshifts $z \simgt 1.5$. In fact our model predicts that about
5\%-10\% of the DLAs should have metallicities of $\rm [M/H]<-3$ in
the redshift range $z=2-4$. In contrast there is an apparent observed
`floor' in DLA metallicities. In particular, in the sample analysed by
\citet{2003ApJ...595L...9P}  no single DLA system was observed
with a metallicity $\rm [M/H]\simlt -3$ . These authors use
$\alpha$-elements and Fe measurements, in addition to Zn measurements,
and so are able to detect lower metallicities than would be possible
with Zn alone.  At face value this suggests a discrepancy, however our
metallicity model assumes that the primordial infalling gas has zero
metallicity. By allowing for a mild pre-enrichment of the
intergalactic medium to the level of $\rm [M/H]=-4.5$ the
$99\%$-contour would move to $\rm [M/H]\sim -4$ with the
$95\%$-contour also at higher metallicity resulting in better
agreement with the apparent observed metallicity floor. We will return
to this issue in Section 3.3, where we discuss possible  effects of
selection biases and dust obscuration.

Comparing to the observations, we see that the majority of the data
points lie within the $75\%$-contour with some outliers at high
metallicities. Especially at around $z=2$ a few DLAs have been
observed with `anomalously' high metallicities of almost solar value. In
fact, the highest metallicity system observed by
\citet{2002A&A...392..781L} has $\rm [M/H]=-0.11$. These authors argue
that this DLA is probably connected to a star formation region because
they find evidence of dust and a relatively high molecular hydrogen
fraction. Our model does not include spatial inhomogeneities within
galaxies which could affect the predicted metallicity
distributions. Furthermore our model does not include any variance in
the strength of stellar feedback, which as we will show in the next
subsection could also influence the predicted distributions.

In summary the results of this Section show that our model reproduces
the lack of evolution in the metallicity of the DLA systems seen in
the observations over the wide redshift range $z\approx 4$ to
$z=0$. In addition, the mean metallicity and dispersion predicted by
the model also provides a good match to the observations. There may be
some discrepancies between the predicted metallicity distributions and
the observations at both extremely high and low metallicities.
However as we will argue later in this Section the tails of these
distributions are sensitive to model parameters and selection biases.

\begin{table}
 \label{met_fit_linear}
 \centering
  \begin{tabular}{@{}lll}
  \hline
   Data set & $m$     &  $\rm [M/H]_{0}$   \\ 
   \hline\hline
   NFB    & $-0.04$  & $-0.76$ \\ 
   NFB unweighted & $-0.02$  & $-1.26$ \\  
   WFB  & $+0.03$  & $-0.69$ \\ 
   SFB  & $-0.07$  & $-0.88$ \\ 
   NFB bias-corrected & $-0.03$  & $-0.79$ \\ 
   NFB dust-corrected & $-0.03$  & $-0.84$ \\ 

\hline
\end{tabular}
\caption{The best fit linear fits to the metallicity evolution with redshift for the various
	data sets, $m$ is the slope in units of $\rm dex/\Delta z$ and $\rm [M/H]_{0}$ the zero point in units of dex.
	All metallicities weighted with $N_{\rm HI}$, unless otherwise stated.} 
\end{table}

\begin{figure*}
 \begin{center}
   {\includegraphics[scale=0.8]{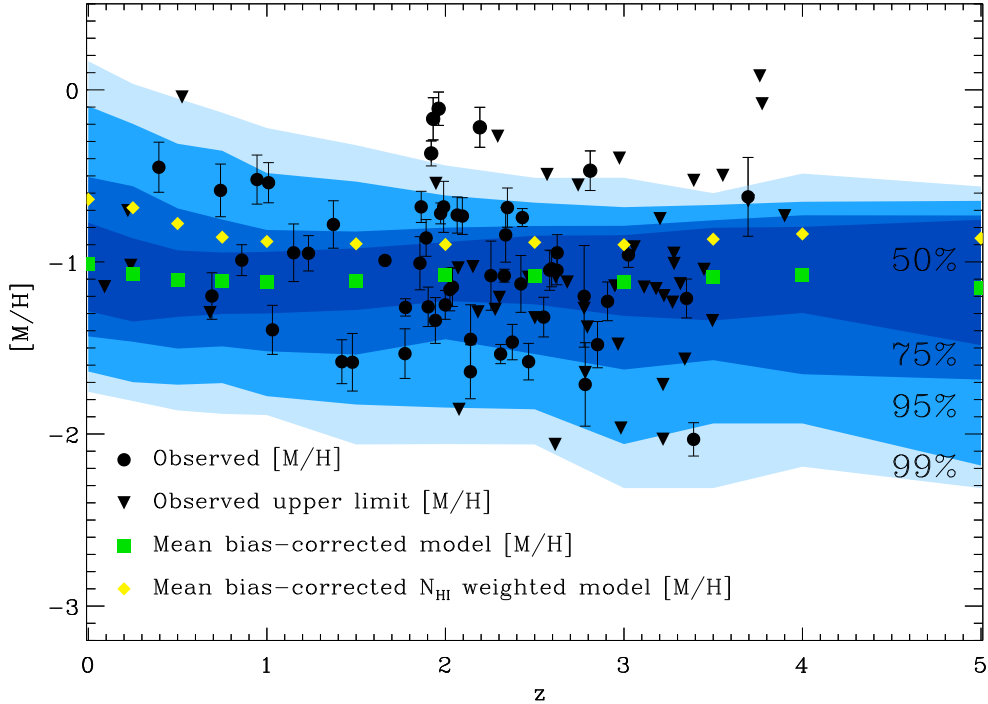}}
 \end{center}
 \caption[]{The redshift evolution of DLA metallicity for the bias-corrected normal feedback model ($\phi_{\kappa}=0.1$) overplotted 
           with the same observational data and symbols as Fig. \ref{fig:Met_redshift_norm}.}
 \label{fig:Met_redshift_bias_norm}
\end{figure*}

\subsection{Dependence on the strength of stellar feedback}

We also calculated the metallicity evolution adopting strong
($\phi_{\kappa}=1.0$) and weak ($\phi_{\kappa}=0.01$) feedback
prescriptions. The predicted metallicity distributions as a function
of redshift are plotted in
Fig. \ref{fig:Met_sfb_wfb_redshift_norm}. As expected, the SFB model
produces strong outflows that eject a high fraction of the metal
enriched gas from the galaxies, especially in their early phases of
evolution. The overall effect is to produce a net metallicity that is
0.1-0.3 dex lower than for the NFB model, with the difference being
larger at high redshifts.  Fitting the linear evolution of the $N_{\rm
HI}$-weighted metallicity we get $m= -0.07 \ \rm dex/\Delta z$ with
$\rm [M/H]_{0}=-0.88 \ \rm dex$. This model produces slightly stronger
evolution in the metallicity because stronger outflows lower the
metallicities at high redshift in comparison to the NFB model.
However, the main effect of the strong feedback is to lower the
overall mean metallicity, making it more difficult to explain the
observed high metallicity DLAs.
 
The WFB model produces mean metallicities that are 0.1-0.3 dex higher
than for the NFB model.  The main differences with the NFB model are at
high redshift where the reduced effects of outflows allow the buildup
of high mean weighted metallicities of $\rm [M/H]\sim -0.5$.  This
high early enrichment results in a small positive evolution in
metallicity with increasing redshift $m= +0.03 \ \rm dex/\Delta z$
with $\rm [M/H]_{0}=-0.69 \ \rm dex$.

Both the strong and weak feedback models produce essentially no
evolution in the metallicity.  The main effect of altering the
feedback efficiency is to raise or lower the overall metallicity by a
relatively small amount.  Interestingly our default NFB model comes
closest to providing a match to the observations. Evidently the
assumption of a uniform feedback efficiency is a gross
oversimplification. More realistically, we would expect a range of
feedback efficiencies caused by both differences in the properties
of the interstellar medium in galaxies (cold cloud temperatures, sizes, heating mechanisms
{\it etc.}) and from transient departures from our self-regulated model of star
formation ({\it e.g.} stronger stellar feedback in
interacting systems and weaker feedback in more isolated
systems). Variations in feedback efficiencies might well affect the
overall metallicity distributions of DLA systems particularly in the
tails at high and low metallicities. However, although it would be
possible to construct an ad-hoc model with variable feedback
efficiencies, developing a physically motivated model is beyond the
scope of this paper.

\subsection{Selection biases and dust}

The model predictions of the previous sub-section were
computed  assuming that the effects of dust and other
observational biases are negligible.
\citet{1998A&A...333..841B} report an anti-correlation between the
observed Zn abundance and $N_{\rm HI}$, independent of redshift.  This
result has been confirmed by \citet{2000MNRAS.315...82P} who note that
all observed DLAs are found between $18.8 < \rm [Zn/H] +\log (N_{\rm
HI})< 21$ in the $\rm [Zn/H]-N_{\rm HI}$ plane. The lower limit in
this relation may arise because low column density and
low-metallicity systems are not detectable through their absorption
lines in current surveys.  There are several possible explanations for
the lack of DLAs with high column density and high-metallicity: a) The
light of the background quasars may be severely extinguished by dust;
b) the cross-section for high $N_{\rm HI}$ absorption may decrease with
increasing metallicity as gas is consumed by star formation; c) systems 
with $\log (N_{\rm HI})> 21$ may be intrinsically rare
\citep{2001ApJ...562L..95S}.

The left-hand panel of Fig. \ref{fig:Met_Nh_bias} shows our default
$\phi_{\kappa}=0.1$ model predictions in the $\rm [Zn/H]-N_{\rm HI}$
plane. The straight lines show the \citet{2000MNRAS.315...82P} bias
thresholds.  The fraction of our models that lie above the upper
cutoff (and therefore may be biased by dust obscuration) is always low
except for the higher redshifts where its contribution is of order
$7\%-8\%$ at $z=4-5$. However we find that a significant fraction of our
models lie below the lower threshold (typically 10-20\%).  It is
therefore clear that removing points below the lower line will have a
significant effect on metallicity distributions, whereas removing
points above the upper line will have little effect. We have also
overplotted the observational data on this diagram.  At redshifts
$z<1$ the observations match our models reasonably well but at $z>2$
the models fail to match the observational points with $\log N_{\rm
HI}< 20.5$ and metallicities $\rm [Zn/H]>-1$. In the bin centered on
$z=4$ there is almost no overlap between the observations and our
NFB model, though at high redshift all but one of the quasar selected metallicity
observations are upper limits. We have also plotted the recent metallicity determination for
a DLA system at $z=4.048$ detected against the optical after-glow of the gamma-ray burst
GRB060206 \citep{2006A&A...451L..47F}. This lies much closer to the model
predictions.

Recently there has been considerably interest in 
sub-damped Lyman-$\alpha$ systems 
\citep[see {\it e.g.}][]{2003MNRAS.346..209L,2003MNRAS.345..447D} with column densities
of $10^{19} \rm cm^{-2} \leq N_{\rm HI} \leq 2 \times 10^{20} \rm cm^{-2}$, i.e. 
below the classical definition of DLAs. The analysis by \cite{2005MNRAS.363..479P} found that the 
sub-DLAs contribute about 20\% to the total neutral gas mass density at redshifts 
$2\leq z\leq 5$. Furthermore, \cite{2006A&A...450...53P} recently discovered a sub-DLA with
super-solar metallicity at $z=0.716$ indicating that sub-DLAs might be more metal-rich than classical DLAs. 
We do not consider sub-DLAs in the present analysis. However, we note that a large number of sub-DLAs with 
low-column densities and high metallicities cannot be explained within the framework of our model as discussed below.

Our model cannot produce large numbers of low column density systems
at high redshift with high metallicity.  As can be seen from
Fig. \ref{fig:Rad_gasmet_prof} the model tends to produce flat $N_{\rm
HI}$ distributions of high column density with a sharp
cutoff. Clearly, the sharpness of the cut-offs are an idealization and
a consequence of applying a simplistic model of self-regulated star
formation based on the stability criterion of equation
(\ref{eq:stability}). In reality, there are many physical process that
will tend to `smear' the cut-offs. These include tidal disruption and
realistic deviations from axial symmetry both during collapse, and
from disc instabilities. (In this context, it is worth noting that the
highest column densities within the Magellanic Stream are only just
below the DLA column density threshold \citep{2003ApJ...586..170P},
suggesting that tidal debris may contribute to the DLA cross section),
Additional effects, not incorporated in our model, that might increase
the spread in column densities at high metallicities include bursts of star
formation (triggered by, say, an interaction) that can consume much of
the gaseous disc, or ram pressure stripping of disc gas. We note that
the numerical simulations of \citet{2004MNRAS.348..435N} produce a
wide spread of column densities extending down to the DLA 
threshold. However, they find a strong correlation between metallicity
and column density and (as in our models) cannot account for the low
metallicity, high column density, systems observed at high redshift.
Understanding the column density distributions plotted in
Fig. \ref{fig:Met_Nh_bias} in greater detail poses an interesting
challenge for both theory and observations.

Our simplified model for the age distribution of disc galaxies can
affect some aspects of our model at high redshift. This is discussed
in greater detail in Section 4, but to illustrate the effects on the
column density distributions we plot in Fig. \ref{fig:Met_Nh_bias} the
evolution of a  model with coeval evolution and a formation
redshift set to infinity (model NFBc) in addition to our standard
models using the \citet{1993MNRAS.262..627L} (hereafter LC93)
 age distributions. In the
coeval model, discs have more time to evolve, especially at high
redshifts as discussed in Section \ref{sec:globprop}. The additional
evolution time shifts the $N_{\rm HI}$ distributions towards lower
column densities, (and the gas metallicities to slightly lower values)
but not by enough to match the observations at $z \simgt 3$.

In Fig. \ref{fig:Met_redshift_bias_norm} we plot the metallicity
evolution with redshift for the DLAs in our models that lie within the
diagonal lines plotted in Fig. \ref{fig:Met_Nh_bias}, {\it i.e.} the
relations that according to \citet{2000MNRAS.315...82P} roughly
delineate regions in the diagram that might be affected by 
observational selection biases.  Including the lower limit,
$18.8 < \rm [Zn/H] +\log (N_{\rm HI})$, makes quite a large difference
because it eliminates most of the systems with metallicities less than
$\rm [M/H]=-2$.  After this bias-correction almost all of the
observations at low metallicity lie within the 99\% contour predicted
by our model. It may therefore be necessary to understand a bias of
this type in more detail to interpret the Zn abundances at very low
metallicities. This type of bias will be less important in direct
observations of Fe, which is more abundant than Zn. Indeed as we
have mentioned previously \citep{2003ApJ...595L...9P} find abundances
as low as $\rm [M/H]\sim-3$ using Fe and $\alpha$ elements.

The bias-correction at high column densities removes some DLA systems
in our models with high metallicity, which makes it even more
difficult to explain the observed high metallicity systems at $z \sim
2$.  Fitting a linear evolution model to the `bias-corrected'
model predictions of Fig.  \ref{fig:Met_redshift_bias_norm} 
 results in $m= -0.03 \rm dex/\Delta z$ with $\rm
[M/H]_{0}=-0.79 \ \rm dex$, i.e. virtually identical to fitting the
uncorrected results shown in Fig. \ref{fig:Met_redshift_norm}.

If we only impose the upper cut, $\rm [Zn/H] +\log (N_{\rm HI})> 21$,
the fit to the metallicity evolution gives $m= -0.03\ \rm dex/\Delta
z$ with $\rm [M/H]_{0}=-0.84 \ \rm dex$. In this case our model
predicts a slightly lower metallicity of $\sim0.1 \ \rm dex$ compared
to the uncorrected results of
Fig. \ref{fig:Met_redshift_norm}. According to our model only a small
number of systems are expected to lie above the line defined by the
line $\rm [Zn/H] +\log (N_{\rm HI})> 21$, and removing the ones that
do lie above this line has very little effect on the overall
metallicity distribution.  Dust obscuration
in high metallicity, high column density systems, is therefore unlikely
to lead to significant biases in the DLA metallicity distributions
according to our model, since such  systems are rare. This is
consistent with the results of \citet{2005A&A...440..499A}, who find
only a marginally higher metallicity of 0.2 dex for their unobscured
(radio-selected quasar) CORALS DLA sample compared to a control sample
from \citet{2005ApJ...618...68K}.

\begin{figure*}
 \begin{center}
   {\includegraphics[scale=0.8]{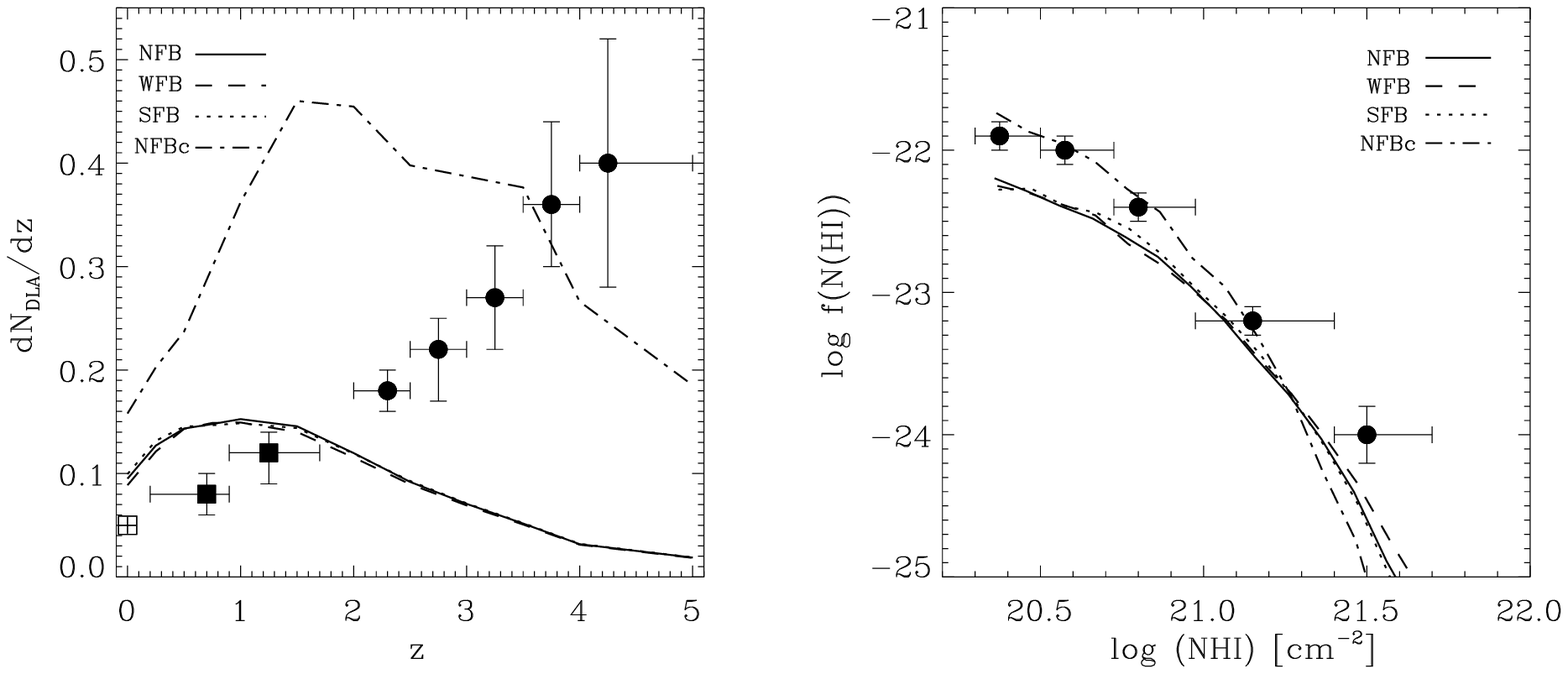}}
 \end{center}
 \caption[]{The DLA rate-of-incidence $dN_{\rm DLA}(z)/dz$ (left-hand
            panel) for the three feedback models and the coeval
            evolution model is plotted with data from
            \citet{2004PASP..116..622P} (solid circles at $z>2$),
            \citet{2006ApJ...636..610R} (solid squares at $z\sim1$)
            and the open square at $z=0$ from
            \citet{2005MNRAS.364.1467Z}.  In the right panel the DLA
            column density distribution $f(N_{\rm HI})$ is shown for
            the four models overplotted with observational data (solid
            symbols) from \citet{2000ApJ...543..552S}. The model
            results are averaged over $0 \leq z \leq 5 $. }
 \label{fig:fnHI_nz}
 \end{figure*}

\begin{figure*}
 \begin{center}
   {\includegraphics[scale=0.8]{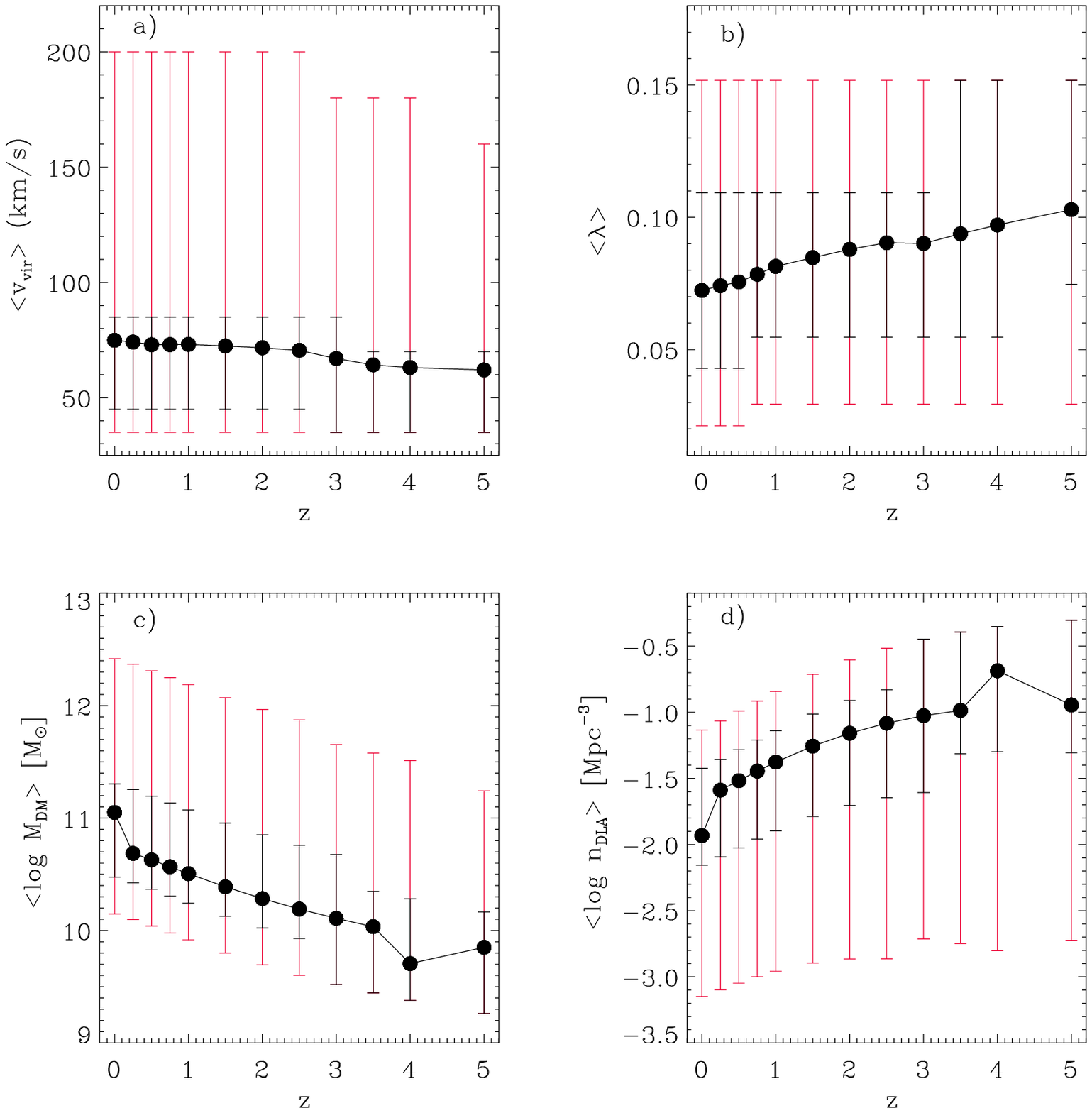}}
   \end{center} \caption[]{The evolution of the mean $v_{\rm vir}$ and
   $\lambda$ with 50\% (quartile, black lines) and 95\% (red lines) ranges (top-panel).  Evolution of
   the median DLA halo mass and comoving halo density with 50\%
   (quartile, black lines) and 95\% (red lines) ranges (bottom-panel). For points where only one
    low/high error bar is shown the 50\% and 95\% ranges are the same, which is a result of the limited
    number of bins in our model (17 in $v_{\rm vir}$, $M_{\rm DM}$ and $n_{\rm DLA}$; 7 in $\lambda$, see 
    Section \ref{sec:Modelling} for details). All plots are for the normal
   feedback model ($\phi_{\kappa}=0.1$).}  
   \label{fig:mean_halo_prop}
   \end{figure*}

\section{Global properties of DLAs}
\label{sec:globprop}

In this Section we compute various global properties of the DLA
population.  Many of the properties discussed in this Section are
sensitive to the merger history of the absorbers, and hence to the age
of the model galaxy that determines its cross-section and mass. As
explained in Section 2.2, we have employed a simplified description
based on the LC93 model for the age distributions of dark matter
haloes as a function of their virial velocity. The high density
baryonic systems within these haloes will be longer lived than the
haloes themselves and will follow a more complex merger history. The
metallicity distributions described in the previous Section are
insensitive to models of the merger history because the metallicity of
the gas is not a strong function of age.  Likewise, the gas
metallicity gradients in our models are shallow and so the metallicity
distributions are insensitive to any process, such as tidal stripping,
that might truncate the gaseous discs.  The metallicity distributions
discussed in the previous Section are therefore robust and are largely
a consequence of the model for feedback and self-regulated star
formation.

The cross-sections, on the other hand, are extremely sensitive to the
age distribution and hence to the merger history. Although it is
possible, in principle, to construct a more elaborate analytic merger
model for the baryonic components \citep[see {\it
e.g.}][]{2005MNRAS...364..997P} the physics involved is complex and
difficult to model accurately.  To illustrate the sensitivity of
various results to the merger history, in addition to our standard
models using the LC93 age distributions, we have therefore calculated
the evolution of a model with coeval evolution and a formation
redshift set to infinity (model NFBc). The NFB and NFBc models provide
two extreme representations of the merger history of early disc
systems that should bracket the predictions of a more realistic model.
The differences between the two representations is particularly
important at high redshift.  At $z=5$ the age of the coeval population
is 1.2 Gyrs compared to the LC93 model which predicts ages for the
models of only 0.3 Gyrs. This difference of a factor of four leads to
large differences in the calculation of cross-sections, redshift
number densities and $\Omega_{\rm HI}$.

To summarize, our model provides robust predictions for the
metallicity evolution of DLAs. Properties that are sensitive to the
disc age distributions and cross-sections, particularly the
rates-of-incidence (see equation \ref{eq:rate-of-inc}) and (to a
lesser extent) $\Omega_{\rm HI}$ are less reliably predicted by
our model, especially at high redshifts. Many of the results presented
in this Section are therefore meant to be indicative at the
`order-of-magnitude' level, rather than detailed fits to the
observations.  To make accurate predictions of such properties, it is
likely that high resolution numerical simulations that properly model
the formation, evolution and merger history of dwarf galaxies will be
required \citep[\ {\it cf.} ][]{2004MNRAS.348..421N}.

\subsection{Abundance of DLAs as a function of column density and redshift}

Following \citet{2004MNRAS.348..421N} (see also \citealt{2005ARA&A..43..861W} and references therein)
we can compute the total DLA rate-of-incidence from
\begin{equation}
\frac{dN_{\rm DLA}}{dz}(z)=\frac{dr}{dz} \int_{M_{\rm min}}^{\infty}
n_{\rm dm}(M',z)\sigma_{\rm DLA}(M',z)dM',
\label{eq:rate-of-inc} 
\end{equation}
where $n_{\rm dm}(M',z)$ is the \citet{1999MNRAS.308..119S} (comoving) dark
matter halo mass function, $\sigma_{\rm DLA}(M',z)$ is the comoving
DLA cross-section and $dr/dz=c/H(z)$, with
$H(z)=H_{0}\sqrt{\Omega_{m}(1+z)^{3}+\Omega_{\Lambda}}$.
The results for our models are plotted in the left panel of
Fig. \ref{fig:fnHI_nz} together with observations from
\citep[]{2004PASP..116..622P,2005MNRAS.364.1467Z,2006ApJ...636..610R}.

We discuss first the results for our standard models based on the LC93
age distributions.  The strength of feedback has almost no effect on
the rate-of-incidence, but the overall shape is very different from
the observations. At redshifts below two the models match the
observations reasonably well (within a factor of two) but at higher
redshifts, the models significantly underproduce $dN_{\rm DLA}/dz$ by a
factor of five. The main reason for the underproduction of
$dN_{\rm DLA}/dz$ at high redshifts is that the prescription of LC93 leads
to galaxies of very young ages (and therefore small cross-sections) at
high redshifts. Assuming coeval evolution as described above (model
NFBc) produces $dN_{\rm DLA}/dz$ which is higher by a factor of five at
high redshifts, roughly matching the observational data at $z \simgt
4$. However, the coevolution model significantly overpredicts
$dN_{\rm DLA}/dz$ at lower redshifts.

As we have argued above, the true age range of the galaxy distribution
is likely to lie somewhere between the LC93 and coeval distributions
and so Fig. \ref{fig:fnHI_nz} suggests that it may be possible to
account for the observed rate-of-incidence with a model of the type
described here incorporating a more realistic merger history. The
numerical simulations of \citet{2004MNRAS.348..421N} actually
reproduce the observed $dN_{\rm DLA}/dz$ quite well and in fact their
comoving tabulated cross-sections as a function of $v_{\rm vir}$ lie
in between the cross-sections of our NFB and NFBc models. It would be
interesting to make a more detailed comparison of our models with age
distributions and cross-sections from numerical simulations to
determine the key physics required to reproduce the observed
rate-of-incidence.

The right-hand panel of Fig. \ref{fig:fnHI_nz} shows the differential
distribution of $f(N_{\rm HI},X(z))=d^{2}N/dN_{\rm HI}dX$ predicted by
our models.  This quantity gives the number of DLAs per unit column
density $N_{\rm HI}$ and per unit absorption distance
\citep{1969ApJ...156L...7B}.  The data points plotted on the
right-hand panel of Fig. \ref{fig:fnHI_nz} are from
\citet{2000ApJ...543..552S} adjusted for our $\Lambda$CDM cosmology.
The model results are calculated by averaging $f(N_{\rm HI})$ over the
redshifts $0 \leq z \leq 5 $ and all three models fit the
observational data reasonably well at high column densities but
underproduce  systems at the low column density end by a factor
of two. The three models differ only at the highest column densities,
where the weak feedback  model produces slightly more high
column density systems compared to the normal and strong feedback
models. On the other hand, the coeval evolution model fits the
observational data nicely with the exception of the highest column
density point. Again, as with the rate-of-incidence analysis, the two
extreme merger models roughly bracket the observational data.

\begin{figure*}
 \begin{center}
   {\includegraphics[scale=0.8]{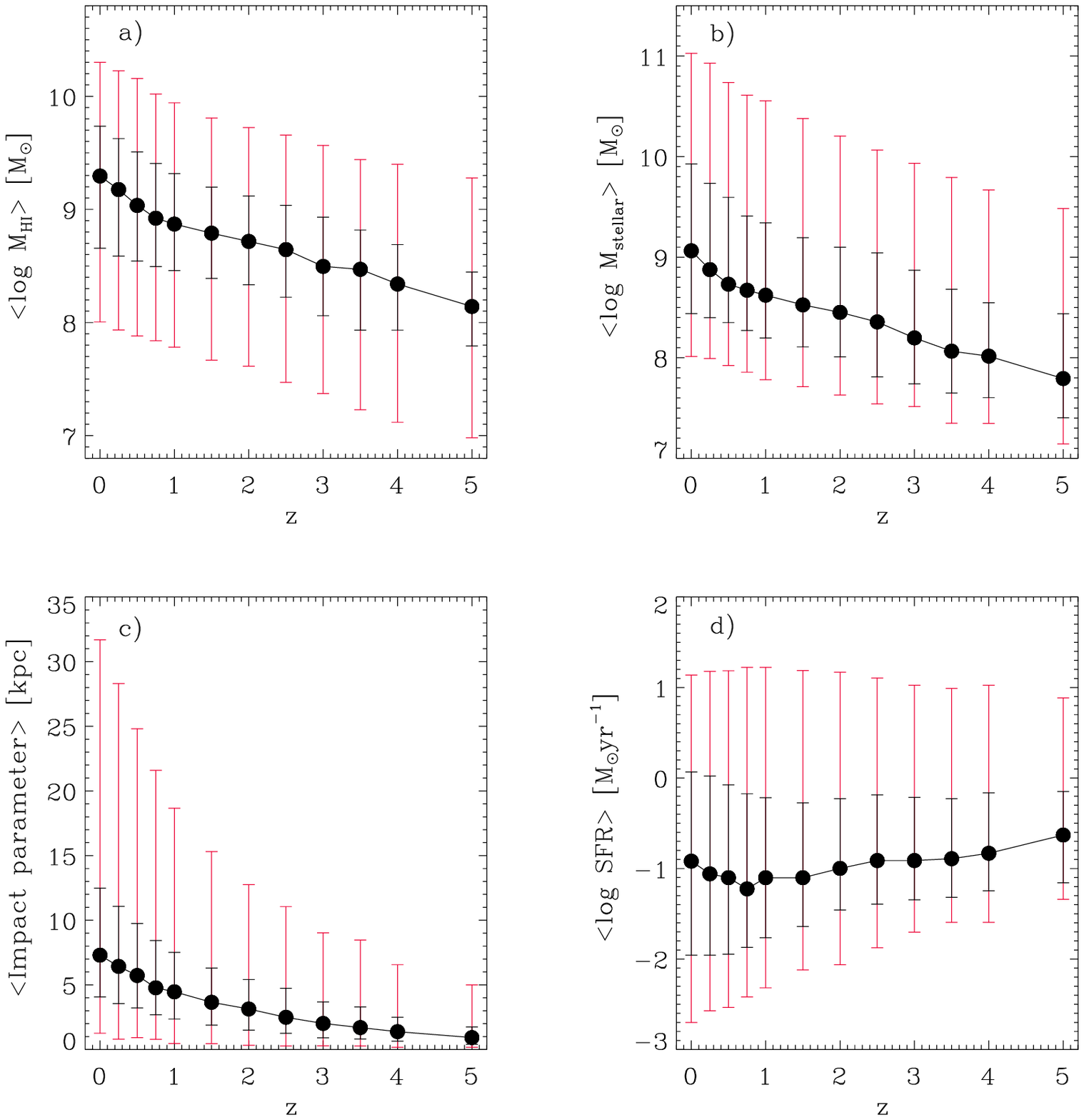}}
 \end{center} 
 \caption[]{The evolution of the median DLA HI mass, median stellar
            mass, median impact parameter and median star formation
            rate with 50\% (quartile, black lines) and 95\% (red lines) ranges.  All plots are for
            the normal feedback model ($\phi_{\kappa}=0.1$).}
 \label{fig:median_gal_prop}
 \end{figure*}

\subsection{DLA halo properties}

In the two next sub-sections we discuss the properties of typical
DLA haloes and of the galaxies that lie at their centres. Unless otherwise
stated, we show results for NFB model using the
LC93 halo age distributions.

The evolution of the average virial velocity of the DLA haloes,
together with the 50\% (quartile, black lines) and 95\% (red lines) ranges, 
are shown in Fig. \ref{fig:mean_halo_prop}a. 
We find a gradual but mild
increase of the average velocity from $\bar{v}_{\rm vir}\sim 60 \ \rm
kms^{-1}$ to $\bar{v}_{\rm vir}\sim 75 \ \rm kms^{-1}$ with a large scatter.
The contribution of
systems with $v_{\rm vir}\geq 100\ \rm kms^{-1}$ to the DLA
cross-section increases from 14\% at $z=5$ to 22\% by $z=0$. The
contribution of even higher virial velocities of $v_{\rm vir}\geq 200\
\rm kms^{-1}$ ({\it i.e.} haloes hosting large Milky-Way type galaxies)
remains low at less than 2\% at all redshifts.  
 Our results agree well with those from the semi-analytic  models of
\citet{2004ApJ...603...12O}, especially at high redshift. At lower
redshift, the typical virial velocities are
somewhat lower than theirs. However our  results for $z \approx 0$ are
 in better agreement with \citet{2005MNRAS.364.1467Z} who infer a mean virial
velocity for the local DLA population of $\bar{v}_{\rm vir}\sim 70 \; \rm
kms^{-1}$.

In Fig. \ref{fig:mean_halo_prop}c we show the
corresponding evolution in the median mass (within the virial radius)
of the dark matter haloes containing DLAs. We find that the median
mass increases from $M_{\rm DM} \sim 5 \times 10^{9} M_{\odot}$ at
$z=4-5$ to a few times $10^{10} M_{\odot}$ at $z=0.5-3.5$ with an
upturn to $M_{\rm DM} \sim 10^{11} M_{\odot}$ at $z = 0$.  Our halo
masses are somewhat lower than those inferred by
\citet{2004MNRAS.354L..25B} in an observational study of MgII
absorbers in the redshift range $0.4 \le z \le 0.8$.  Their results
imply that the hosts of MgII absorbers (typically 40\% of MgII
absorbers are DLAs) have halo masses of $\sim 2-8 \times 10^{11}
M_{\odot}$, perhaps suggesting that MgII selection biases samples
towards higher virial velocities than the average. The abrupt increase
at $z \simlt 0.25$ is caused by the exhaustion of gas in low mass haloes which
placing them below the DLA selection criterion of
$N_{\rm HI} \geq 2 \times 10^{20} \rm cm^{-2}$.

We conclude that the DLA cross-section is dominated at all redshifts
by systems residing in small mass (low virial velocity) haloes with a
mild evolution towards systems with higher mass at low redshift. This
is in agreement with recent semianalytic \citep{2004ApJ...603...12O}
and numerical investigations \citep{2003ApJ...598..741C} but in strong
disagreement with the classical picture in which the majority of DLAs at
all redshifts are massive disc galaxies like our own
\citep{1986ApJS...61..249W}.

The spin parameter $\lambda$ of the DLA haloes as a function of
redshift is shown in Fig. \ref{fig:mean_halo_prop}b.  As DLAs are selected by cross-section,
one would expect DLAs to be biased towards bigger discs with larger
$\lambda$ in comparison to the general galaxy population. This is seen
in our models.  The DLA cross-section is dominated by systems in
haloes with large $\lambda$, especially at high redshift where 
high $\lambda$ systems rapidly build up large gaseous discs. Systems
with small $\lambda$ also experience more extensive mass loss caused by the
feedback from supernovae (see Fig. \ref{fig:Eject}).  At 
lower redshifts the mean cross-section weighted $\lambda$
decreases.  This is explained by the slower buildup of  gaseous discs in 
lower $\lambda$ systems. In summary, we conclude that the 
DLA cross-section is dominated by systems with larger spin parameters 
than the mean, and that the bias is accentuated at high redshifts.

The comoving number density of the DLA parent haloes is shown in
Fig. \ref{fig:mean_halo_prop}d. Mirroring the evolution of the typical
DLA halo mass with $n_{\rm DM} \sim 0.1 \ \rm Mpc^{-3}$ at $z=4-5$
decreasing to $n_{\rm DM} \sim 0.01 \ \rm Mpc^{-3}$ by z=0.  This shows
that the DLA population acquires  an increasing contribution at lower
redshifts from more massive haloes with lower comoving number
densities. We can see a jump in both the masses and number densities
between $z=5$ and $z=4$, this is again caused by the slower buildup of
the smallest mass systems, which only start to fully contribute to the
DLA cross-section at $z \approx 4$.

\subsection{DLA galaxy properties}

Fig. \ref{fig:median_gal_prop} shows the evolution of various
properties of the DLA galaxy population. The DLA median hydrogen gas
mass evolves by a factor of ten from $M_{\rm HI}=10^{8} M_{\odot}$ at
$z=5$ to $M_{\rm HI}=2 \times 10^{9} M_{\odot}$ at $z=0$. Our model
prediction at $z = 0$ is in excellent agreement with
\citet{2005MNRAS.364.1467Z}.  The evolution of $M_{HI}$ is associated
with an increase in the median stellar mass from $M_{\star}\sim 6
\times 10^{7} M_{\odot}$ at $z=5$ to $M_{\star}\sim 10^{9} M_{\odot}$
by $z=0$. According to our model, the DLA cross-section is dominated
by `dwarf' (low-mass) galaxies with large HI discs at all redshifts
and especially at higher redshifts. However, the width of the
distribution (indicated in the figure) increases at lower redshifts
because higher mass systems make a growing contribution to the DLA
cross-section. The contribution of Milky Way-type systems, with total
disc masses of $M \geq 5 \times 10^{10} M_{\odot}$, to the
cross-section is completely negligible at high redshifts, growing to
only 0.5\% at $z=2$.  At lower redshifts this fraction grows to 2\% at
$z=1$ with an increase to 6\% by the present epoch. These results
agree well with the semi-analytic models of
\citet{2004ApJ...603...12O}, who found a $\sim5\%-10\%$ contribution
of Milky Way type systems at low redshift with negligible contribution
to the cross-section at $z \geq 3$. According to our model, the
majority of the DLA population must consist of `dwarf' galaxies with
typical masses of $\simlt 0.1 M^{*}$. This result is also consistent
with the lack of direct detections \citep[see {\it e.g.}][]{2005ARA&A..43..861W}.

Figure \ref{fig:median_gal_prop}c shows the
evolution of the median impact parameter (in physical units) as
a function of redshift. We find a very strong, almost linear evolution
in this parameter, from very small values of $\sim 1\; {\rm kpc}$ at
$z=5$ to values of $\sim 7 \ \rm kpc$ at the present epoch. The local
value again agrees  well with the observational results of
\citet{2005MNRAS.364.1467Z} who find a median value for the local
impact parameter of 7.8 kpc. The small impact parameters in
combination with the relatively low number densities of the DLA haloes
at high redshifts results in  the  low rate-of-incidence at high $z$
discussed in Section 4.1 (Fig. \ref{fig:fnHI_nz}).


 \begin{figure*} \begin{center}
 {\includegraphics[scale=0.8]{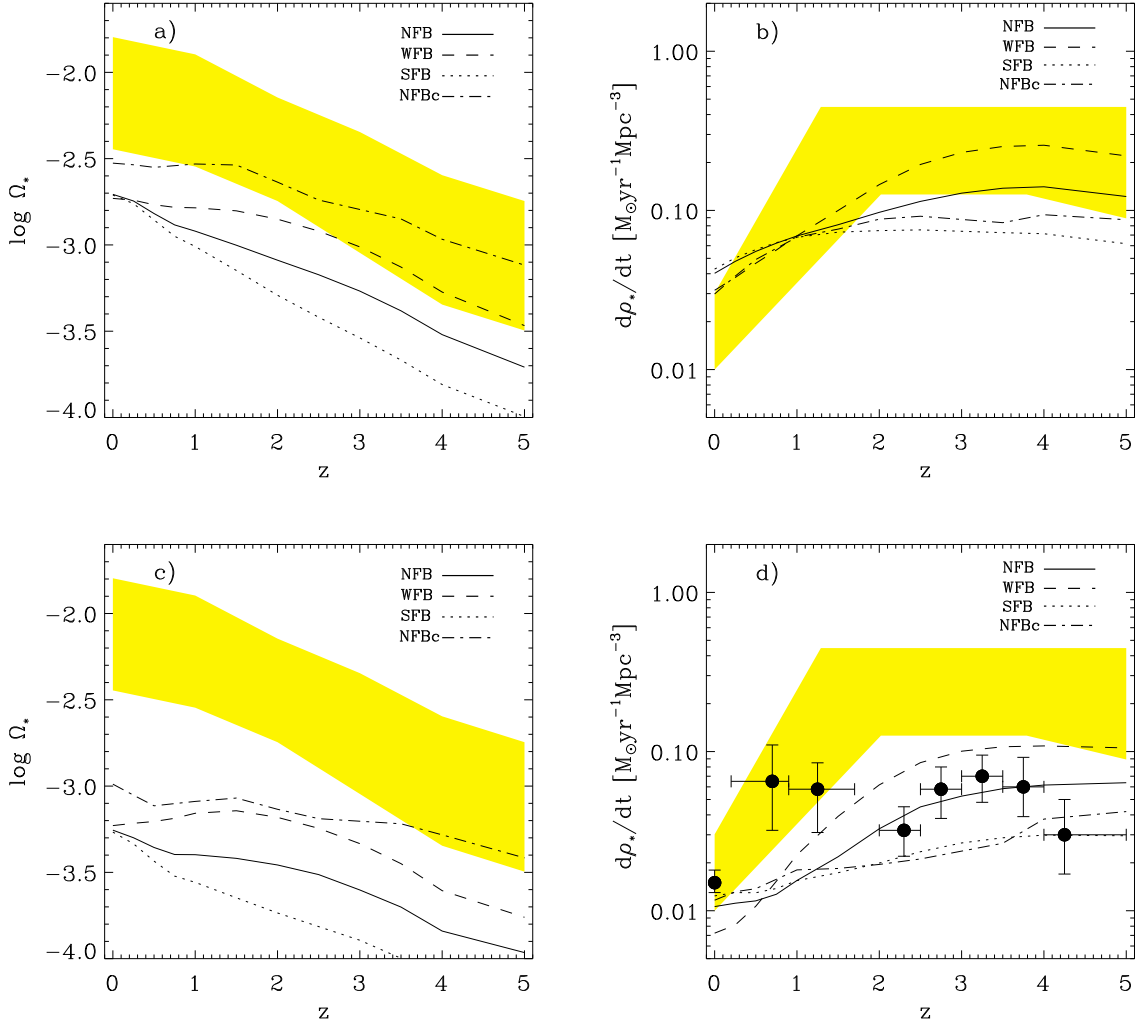}}
 \end{center} \caption[]{The top left panel (a) shows the evolution of
 global volume weighted cosmic star density.  Panel (b) shows the
 evolution of the global volume weighted star formation density. The
 lower panels (c) and (d) show the same quantities but weighted by
 cross-section (above the DLA threshold).  In each case we plot
 results for the three feedback models using the LC93 age
 distributions and for the coeval evolution model with the normal
 feedback model. The shaded regions in the figures show the volume
 averaged stellar densities and star formation rates from
 \citet{2004ApJ...615..209H} and \citet{2005ApJ...630..108H} that
 encompass the majority of the observational points on the
 `Madau'-diagram after correction for dust obscuration. The points 
plotted in panel (d) show the star formation rates inferred by
 \citet{2005ApJ...630..108H} for DLA systems (see text for details).}

\label{fig:Omega_star_SFR_comb}
\end{figure*}

\subsection{Star formation of DLAs}

The typical star formation rates of DLAs according to our model are
shown in Fig. \ref{fig:median_gal_prop}d.  We find that the DLAs have
a very broad distribution of star formation rates, ranging from $\sim
5\times 10^{-3}$ up to $\sim 50$ $M_{\odot}\rm yr^{-1}$ at all redshifts. The
evolution of the median star formation rate is virtually constant with
a median star formation rate of $\dot{M}_{\star}\sim 0.1-0.2 M_{\odot}\rm yr^{-1}$
over the entire redshift range $z=0$--$5$. The mean star formation
rate is considerably higher at $\dot{M}_{\star}\sim 1-2 M_{\odot}\rm
yr^{-1}$. The distribution of star formation rates has a long tail
extending to high values for sightlines probing the central regions of
massive galaxies (see the 95\% ranges in Fig. \ref{fig:median_gal_prop}d).

\begin{figure*}
 \begin{center}
   {\includegraphics[scale=0.8]{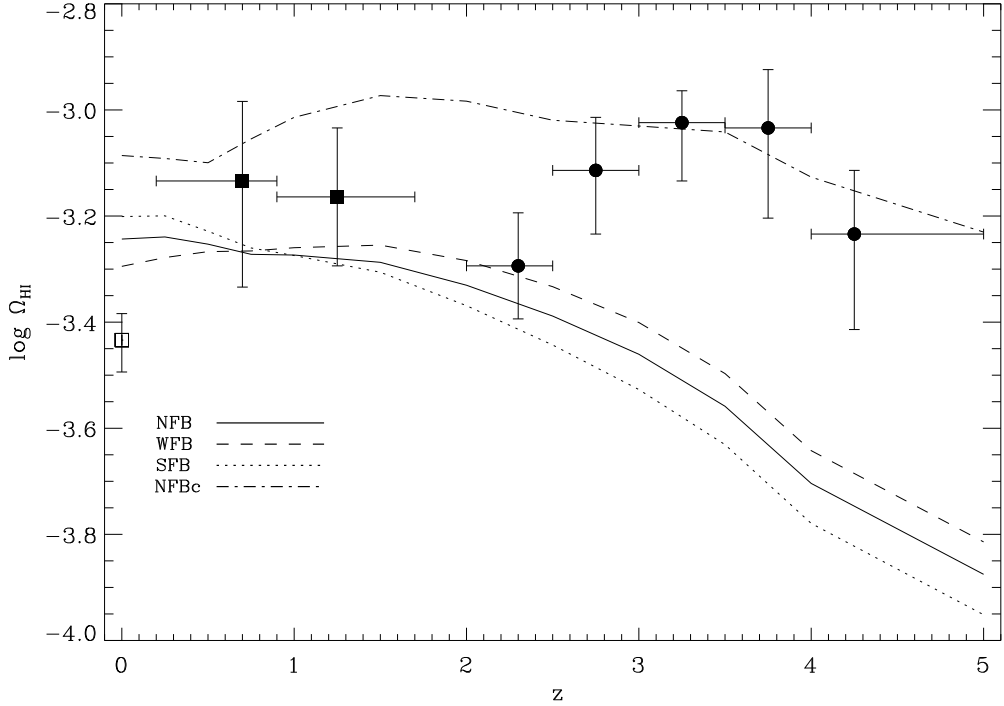}}
   \end{center} \caption[]{Evolution of the cosmic hydrogen density in
   DLAs for our models overplotted with observations
   from \citet{2005MNRAS.359L..30Z} (open square at $z=0$),
   \citet{2006ApJ...636..610R} (solid squares at $z\sim1$) and
   \citet{2004PASP..116..622P} (solid circles at $z>2$). The model
   results have been corrected for a hydrogen fraction of
   $X_{H}=0.7$.}
             
 \label{fig:Omega_HI_comb}
 \end{figure*}

However, we find strong evolution in the star formation rates per unit area, with the
median evolving from $\dot \Sigma_{*}\sim 1.5-3.0 \times 10^{-2}
M_{\odot} \ \rm yr^{-1} \ kpc^{-2}$ at $z=4-5$ to $\dot \Sigma_{*} \sim
10^{-3} M_{\odot} \ \rm yr^{-1} \ kpc^{-2}$ at $z<2$. In our model,
galaxies at high redshift are experiencing bursty star formation (with
high star formation rates per unit area) followed by more quiescent star formation
evolution at late times, when the gas infall rate is regulated by the
Hubble time rather than the free-fall time (see E00 for details).  The
median values of $\dot \Sigma_{*}\sim 10^{-3} - 10^{-2} M_{\odot} \ \rm
yr^{-1} \ kpc^{-2}$ are in good agreement with the semi-analytic model
of \citet{2005ApJ...623...99O} and with the numerical simulations of
\citet{2004MNRAS.348..435N}.

The evolution of the cosmological stellar density and star formation
rates are shown in Fig. \ref{fig:Omega_star_SFR_comb}. We show results
for the three feedback models for the LC93
merger history, and for a model with coeval evolution. It is important
to note that the {\it volume averaged} stellar density and star
formation rates shown in this figure are weighted strongly towards the
inner regions of massive disc systems, which make little contribution
to the DLA cross-sections. As we will show below, these {\it global}
volume averaged properties are largely disconnected from the
properties of the DLA population. It is also worth noting that, unlike
more elaborate semi-analytic models \citep{2006MNRAS.365..11,
2006MNRAS.370..645B}, our model has not been fine-tuned to match these
global properties. Nevertheless, it performs surprisingly well given
its simplicity.

All models show high star formation rates at high redshift arising
from the initial `bursty' phase of star formation model, followed by a
gentle decline in the comoving star formation density.  The
differences between the feedback models are larger at higher
redshifts, where the weaker feedback in the WFB model results in high
star formation rates. At lower redshifts ($z \simlt 2$) all three
feedback models produce star formation rates which are within a factor of two of each
other. The global volume averaged star formation rates are quite
sensitive to the assumed age distributions, especially at high
redshift. 

Fig. \ref{fig:Omega_star_SFR_comb}b show the evolution of the volume
averaged star formation rates in our models. The shaded regions in
Figs 11a and 11b shows the range of observational estimates from the
papers of \citet{2004ApJ...615..209H} and \citet{2005ApJ...630..108H},
including corrections for dust obscuration. Figs 11a and 11b give an
impression of the  uncertainties in both the model predictions and the
observations. The models broadly match the extinction corrected star
formation rates, though it is clear that a more accurate treatment of
the merger history is required to produce reliable predictions.

The main point of this comparison is illustrated in Figs.
\ref{fig:Omega_star_SFR_comb}c and \ref{fig:Omega_star_SFR_comb}d.
These show the evolution of the cosmic stellar density and star
formation rate but now weighted by cross-section above the DLA
threshold, rather than volume averaged over all systems. According to
our model, DLA systems account for a small fraction of the total star
formation at all redshifts. This is broadly consistent with the
results of \citet{2005ApJ...630..108H}, who infer the contribution of
DLAs to $\Omega_*$ and $d \rho_*/dt$ by integrating over the observed
DLA column-density distribution assuming a \citet{1998ARAA..36..189K}
relation between the star formation rate and the gas surface density.
(The \citet{2005ApJ...630..108H} results are shown by the points in
Fig. 11d).  Most of the star formation occurs in the inner parts of
galaxies, whereas the DLA are sampling the outer gaseous discs where
there is relatively little star formation. This, of course, is why the
metallicity distributions are skewed to low metallicities at all
redshifts. One should therefore be extremely cautious in interpreting
models of cosmic chemical evolution that link the volume averaged star
formation rates to the metallicity evolution of DLA selected systems
\citep{1998ApJ...494L..15W,
1999ApJ...522..604P,2005ARA&A..43..861W}. As our models show, these
quantities are linked only indirectly and are very likely probing star
formation in very different environments.

\subsection{Cosmological evolution of $\Omega_{\rm HI}$}

From our models we can calculate the cosmological evolution of neutral
hydrogen. This is an important consistency check of the models, since
unlike the stellar density, most of the {\it volume} averaged density
in neutral hydrogen in our models is contained in DLA systems.
Fig. \ref{fig:Omega_HI_comb} shows observational results from
\citep[]{2005MNRAS.359L..30Z,2006ApJ...636..610R,2004PASP..116..622P}
spanning the entire redshift range $0 \simlt z \simlt 5$. The
observations show little evidence for any evolution in $\Omega_{HI}$
over the redshift range $1 \simlt z \simlt 5$.

Our model predictions are also shown after correcting for a 
hydrogen fraction of $X_{H}=0.7$. No correction has been made for the
fraction of hydrogen in molecular form. The results are relatively
insensitive to the strength of feedback, but are quite sensitive to
the assumed age distributions. The models with the LC93 age
distributions underpredict $\Omega_{\rm HI}$ at $z \simgt 2$, whereas the
coeval model agrees quite well with the observations over the redshift
range $1 \simlt z \simlt 5$. As with several other properties
discussed in this Section, the coeval models with those with the LC93
age distribution roughly bracket the observations.  All of the models
are high (by a factor of $\sim 3$) compared to the observed HI density
at $z=0$ (determined from 21 cm observations
\citep{2005MNRAS.359L..30Z}). We do not regard this discrepancy as
particularly serious, since it can be explained if some of the gas in
the extended outer parts of small disc systems is stripped and
returned to the intracluster or intra-group medium as structure grows.

Most of the neutral hydrogen density in our models is in the outer
parts of dwarf galaxies and plays little role in the volume averaged
star formation history of the Universe. The evolution of $\Omega_{\rm
HI}$ at high redshift is therefore more closely linked to feedback
processes and the spatial distribution of gas in the outer parts of
dwarf galaxies than it is to the cosmic star formation history. Our
models show that the low metallicities, characteristic of DLAs, and
the lack of evolution of $\Omega_{HI}$ are consistent with a picture
in which the DLA sight-lines are preferentially sampling the outer
parts of dwarf galaxies.

\section{Conclusions}
\label{sec:conc}

\noindent

In this paper we have used a model of self-regulated star formation 
and a physically motivated model of stellar feedback to construct a
simple model of damped Lyman-$\alpha$ systems. 

Our models reproduce the low mean metallicities seen in DLAs, the lack
of evolution of the mean metallicity and can account, at least
qualitatively, for the observed spread in metallicities at each
redshift.  In our model, most DLA sight-lines probe the outer gaseous
parts of `dwarf' galaxies ($v_{\rm vir} \simlt 70\; {\rm km}{\rm
s}^{-1}$), where the star formation rates and metal enrichment are
always low. Occasionally, sight-lines intersect the inner regions of
disc systems where the gas metallicities are high. Thus geometry is
primarily responsible for the large spread in metallicities seen in
the observations. The metallicity distributions in our models are
relatively insensitive to the strength of the stellar feedback assumed
and to the assumed merger histories of the galaxies (parameterised
by their age distributions). We therefore believe that the DLA 
metallicities are a robust feature of our model. The metallicities
predicted by our model are lower, and in much better agreement with 
observations, than those found in  numerical simulations
\citep{2003ApJ...598..741C, 2004MNRAS.348..435N}. Differences in the
models for stellar feedback and star formation are the most plausible
reasons for this.

In many respects, our conclusions are similar to those of previous
semi-analytic models \citep{2004ApJ...603...12O} and numerical
simulations  \citep{2003ApJ...598..741C,
2004MNRAS.348..435N}. However, our model differs from previous
semi-analytic calculations in that the infall model, star formation
prescription and feedback model, all key factors in determining the
metallicities and spatial distributions of gas and stars, are all
physically motivated. The model has, therefore, few adjustable
parameters. Indeed, the model was developed specifically to 
investigate the role of stellar feedback during galaxy formation
(E00), yet with no further modification  provides a reasonable
description of the properties of  DLA systems.

Nevertheless, it is clear that the physics behind the DLA population
is complex, and it is important to understand both the strengths and
weaknesses of a simple semi-analytic model of the type described
here. Although the metallicity predictions are robust, the sizes of
the gas discs in our model, particularly at high redshifts, are
sensitive to the assumed age distributions. These distributions, in
turn, depend on the merger history of the gaseous discs. To illustrate
the sensitivity of our results to the age distributions, we have used
two simple models, one based on the age distributions of the dark
haloes (using the prescription of LC93) and the other based on the
assumption that all systems formed at $t=0$ (the coeval model). The
actual age distribution for the baryonic components is likely to lie
between these two extremes.

The rates-of-incidence (Figure 8) are particularly sensitive to the
total gas cross-section, and hence to the model of the age
distribution.  Our model cannot predict this quantity
reliably. Nevertheless, the two models for the age distributions
roughly bracket the observations over the redshift range $1.5 \simlt z
\simlt 5$, suggesting that a more realistic merger model may be able
to account for the data over this redshift range. In fact, the
numerical simulations of \citet{2004MNRAS.348..421N} reproduce the
observed rates-of-incidence quite well, and (not surprisingly) their
cross-sections are intermediate between those of our two
age-distribution models. (The physical mechanisms governing the
cross-sections in the numerical simulations are not clear, however.)
At lower redshift, both age-distribution models overpredict the
rates-of-incidence, suggesting that we are missing some physical
mechanism that can limit the growth of extended gaseous discs at low
redshift. The cosmic density in neutral hydrogen is also dependent,
but to a much lesser degree, on the assumed age distributions. As with
the rates-of-incidence, the two models for the age distribution
roughly bracket the observations, except at $z=0$, where our models
overpredict the local HI density by a factor of $\sim 1.5$--$2.5$.

The relation between the metallicities and column densities is not
well reproduced by our models, particularly at redshifts $z \simgt
2$. In our models, the gaseous discs at any time have roughly constant
surface density and truncate abruptly (with an outer radius fixed by
the angular momentum of infalling gas). As a result, our models fail
to reproduce the spread to low column densities seen in the
observations (though only by a factor of two or so) . The abrupt
truncation is clearly an artificial feature of our models and it is
easy to think of a number of physical mechanisms (discussed in Section
3.2) that would lead to lower column densities in the outer parts of
galaxies,  without altering the metallicity of the gas. The numerical
simulations of \citet{2004MNRAS.348..435N} do show a large range of
column densities, extending below the DLA threshold.  The observed
high proportion of low metallicity, high column density density
($N_{\rm HI} > 10^{21} \; \rm cm^{-2}$) systems at $z \simgt 2$ is
more problematic, both for our model and the numerical simulations.
The resolution of this discrepancy is not at all clear.

Our models can roughly reproduce the observed volume averaged stellar
density, HI density and star formation rate as a function of redshift
(though the models were not finely tuned to do so). The cosmic density
in HI is dominated by gas in DLA systems. However, the stellar density
and star formation rates weighted by cross-section are considerably
lower than their cosmic values over the entire redshift range $0
\simlt z \simlt 5$. In our models, the galaxies responsible for DLAs
make only a small contribution to the total stellar density produced
in the Universe. This conclusion is broadly in line with the results
of \citet{2005ApJ...630..108H}, who used a \citet{1998ARAA..36..189K}
relation to infer the contribution of DLA systems to the volume
averaged star formation rate. According to our model, there is only a
indirect relation between the volume averaged star formation rate,
$\Omega_{\rm HI}$ and DLA metallicities. Thus one should be skeptical
of models of `cosmic' chemical evolution that attempt to relate these
quantities ({\it e.g.} \citet{1998ApJ...494L..15W,
1999ApJ...522..604P,2005ARA&A..43..861W}).

As we have mentioned in the Introduction, one of the main purposes of a
semi-analytic model is to gain physical insight into complex problems.
Ultimately, given high enough spatial resolution, it should be
possible to develop realistic numerical hydrodynamic models of
DLAs. In the meantime, we pose the following  problems for
numerical simulators:

\smallskip

\noindent
(i) In our model, the sizes (and hence cross-sections) of the DLA
systems are set by the high angular momentum gas within the dark
haloes, assuming angular momentum is strictly conserved. (The low
angular momentum gas is preferentially expelled in a wind). Is 
this really true?

\smallskip

\noindent
(ii) We have found that some properties of the DLA population are
sensitive to the age distribution and hence the merger history
of the baryonic systems within dark haloes. What is the actual
merger history of these baryonic systems? To what extent does
this merger history affect the properties of DLAs.

\smallskip

\noindent
(iii) How sensitive are the metallicities of DLAs to star formation
and stellar feedback? Is it possible to incorporate more realistic
models of these processes in numerical simulations? Is it possible to
resolve the discrepancies between observations and simulations
performed so far \citep{2003ApJ...598..741C, 2004MNRAS.348..435N}?

\smallskip

\noindent
(iv) Is is possible to reproduce the observed proportion of low
metallicity systems with high column density, that neither
our model or the numerical simulations seem able to explain?

\smallskip

\noindent
(v) Clearly a theoretical model in which dark haloes are populated by
single circular, homogeneous, gaseous discs is a gross
over-simplification.  To what extent are tidal features and other
irregularities important in understanding the DLA population? How
important are inhomogeneities, such as individual star forming
regions, in explaining the high end of the metallicity
distribution?

\smallskip

\noindent
(vi) Our model is too simplified to model the velocity structure seen in metal-line-systems
associated with DLAs. Can this structure be reproduced by detailed hydrodynamical 
simulations, as suggested by the work by \citet{1998ApJ...495..647H}?

\smallskip

\noindent
(vii) In our model, the DLAs make a sub-dominant contribution to the
cosmic star formation rate at all redshifts. Is this really true?

\smallskip

No doubt the reader can think of many other problems. Despite the
compelling case for further work, we do believe that the simple model
presented here gives a plausible explanation for the observed
metallicities of DLAs, and that these metallicities can only be
understood if the DLAs are predominantly `dwarf'-like systems that
contribute little to the net cosmic stellar density. The `classical'
picture \citep{1986ApJS...61..249W} of DLAs as giant spiral discs
slowly converting their gas into stars seems to us to be
irreconcilable with the observations. Furthermore,  we now know
enough about the primordial cosmological fluctuations, from Mpc scales
to the scale of the Hubble radius \citep{2003ApJS..148..175S,
2005MNRAS.362..505C, 2004MNRAS.355L..23V,2005PhRvD..71j3515S} that it is difficult to
imagine a cosmogonic context for the classical picture.

\section*{Acknowledgments}

We thank Max Pettini and Emma Ryan-Weber for helpful comments and
interesting discussions. We also thank Chris Akerman for providing us
with the observational DLA data. This work has been supported by the
Particle Physics and Astronomy Research Council.

\label{lastpage}

\end{document}